\documentclass[prb]{revtex4}
\usepackage{wrapfig}
\usepackage{amsfonts}
\usepackage{amssymb}
\usepackage{psfrag}
\usepackage{graphicx}
\usepackage{amsmath}
\usepackage{bm}

\begin{document}

\title[Entangled Vortices]{Entangled Vortices: \\ Onsager's geometrical
picture of superfluid phase transitions\footnote{Write-up of lectures
presented at the Kevo Winter School, Kevo, Finland, 20-26 April 2002.}}
\author{Adriaan M. J. Schakel} 
\email{schakel@boojum.hut.fi}
\affiliation{Low Temperature Laboratory, Helsinki University of
Technology, P.O. Box 2200, FIN-02015 HUT, Finland}

\date{June 21, 2002}

\begin{abstract}
Superfluid phase transitions are discussed from a geometrical
perspective as envisaged by Onsager.  The approach focuses on vortex
loops which close to the critical temperature form a fluctuating vortex
tangle.  As the transition is approached, vortex lines proliferate and
thereby disorder the superfluid state, so that the system reverts to the
normal state.  It is shown in detail that loop proliferation can be
described in exactly the same way as cluster percolation.  Picturing
vortex loops as worldlines of bosons, with one of the spatial
coordinates interpreted as the time coordinate, a quantitative
description of vortex loops can be given.  Applying a rotation (to
superfluids) or a magnetic field (to superconductors), which suppresses
the formation of vortex loops and instead can lead to open vortex lines
along the field direction, is shown to be equivalent to taking the
nonrelativistic limit.  The nonrelativistic theory is the one often used
to study vortex lattice melting and to describe the resulting entangled
vortex liquid.
\end{abstract}


\maketitle

\section{Preface}
As classical fluids and gasses, superfluids, Bose-Einstein condensates,
and also superconductors support vortices \cite{kevo}: extended
line-like objects around which the fluid or gas executes a circular
motion.  In the classical world, vortices are better known as whirlpools
(in fluids) and as whirlwinds (in the atmosphere)---or also as tornadoes,
or typhoons, depending on which side of the Pacific Ocean one lives.
The velocity of the circular motion around a vortex line varies as the
inverse of the distance to the rotation axis.  The strong winds close to
the center of a whirlwind---the so-called eye of a tornado---are,
incidentally, responsible for most of the damages caused by typhoons.

A striking difference with classical systems is that the circulation
around these extended line objects in quantum systems is quantized.
Given that a vortex usually is macroscopic in size and sets all the
atoms or electrons of the system in motion, the quantization of
circulation is truly remarkable and cannot in general be related to the
quantization rules of quantum mechanics.  Instead, as is well known, the
quantization here is provided by topology \cite{Coleman}.

The presence of vortices in superfluids and Bose-Einstein condensates is
an emergent property, associated with the spontaneous breaking of global
gauge symmetry \cite{PWA}.  In supercondutors, this symmetry becomes a
local one, which strictly speaking cannot be spontaneously broken, but
the following argument is not affected by it.  The vacuum manifold,
defining the system's minimum of energy, is a circle S$_\mathrm{vac}^1$
parametrized by the phase of the spontaneously broken U(1) gauge
symmetry.  If a vortex is circled once, the phase moves around the
circle S$_\mathrm{vac}^1$ representing the vacuum manifold.  A vortex
therefore provides a map from the circle S$_x^1$ in real space to
S$_\mathrm{vac}^1$.  Such maps are characterized by an {\it integer}
winding number telling how often the map wraps around the circle
S$_\mathrm{vac}^1$ when S$_x^1$ in real space is circled once.  The
integer nature of the winding number explains the quantization of
circulation around vortices in quantum systems.  A vortex of a given
winding number cannot be continuously deformed into one of a different
winding number and hence is topologically stable.

\begin{figure}
\begin{center}
\includegraphics[width=9cm]{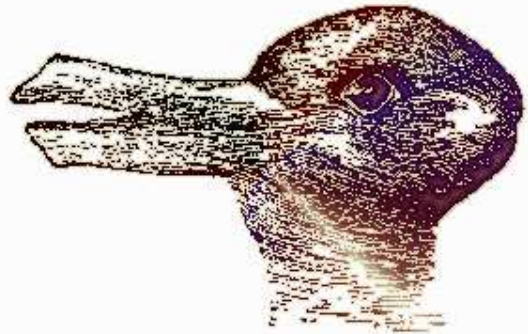}
\end{center}
\caption{Drawing, introduced by the psychologist J.\ Jastrow (1900), that
depicts both a duck and a rabbit.
\label{fig:duck_rabbit}}
\end{figure}

In these lecture notes, superfluid phase transitions are discussed from
the perspective of quantized vortices in the spirit of Onsager's picture
of the $\lambda$ transition in liquid $^4$He \cite{Onsager}.  The
description envisaged by Onsager is one entirely in terms of these
one-dimensional objects, with their geometrical properties such as
fractal dimension and configurational entropy.  In the absence of an
applied rotation (for superfluids) or magnetic field (for
superconductors), vortices cannot terminate inside the system and must
form closed loops.  A superfluid phase transition is characterized in
this picture by a fundamental change in the typical loop size.  Whereas
in the superfluid phase only finite loops are present, at the critical
point infinite loops appear---similar to the sudden appearance of a
percolating cluster in percolation phenomena at critically
\cite{Vachaspati,Bradleyetal,HindmarshStrobl,Akao}.  In Feynman's words
\cite{Feynman55}: ``The superfluid is pierced through and through with
vortex line.  We are describing the disorder of Helium~I.''

The similarity between vortex loop proliferation and cluster percolation
is, as will be discussed, very close.  It derives from a similarity in
the cluster size and loop size distribution.  Both have the same form
containing two factors, one related to the entropy of a given cluster or
loop configuration, and the other related to the Boltzmann weight
assigned to the configuration.  Close to criticality, both of these
factors are parametrized by a single exponent.  One specifies the
algebraic behavior of the distribution at criticality, while the other
describes how the Boltzmann factor tends to unity upon approaching the
critical point.  Clusters and loops that are otherwise exponentially
suppressed can grow without bound when the Boltzmann factor becomes
unity and, thus, gain configurational entropy without energy cost.

The picture of the $\lambda$ transition in liquid $^4$He in terms of
proliferating vortices, which disorder the system and thereby restore
the spontaneously broken U(1) symmetry, has been advanced since then by
various authors \cite{70ies}.  Early numerical evidence for this picture
based on Monte Carlo simulations of the three-dimensional XY-model was
given in Refs.~\onlinecite{Janke,KSW,GFCM}.  Analytic methods to
describe the transition using vortex loops were further developed by
Williams \cite{Williams}, and by Shenoy and collaborators
\cite{Shenoyetal}.  More recent numerical work on the three-dimensional
XY-model from the perspective of vortices, in particular their loop size
distribution, can be found in
Refs.~\onlinecite{AntunesBettencourt,NguyenSudbo,Kajantieetal,Olsson}.

The famous drawing (see Fig.~\ref{fig:duck_rabbit}) introduced by the
psychologist J.\ Jastrow serves as a metaphor for the approach followed
in these notes.  The drawing---a three-dimensional snow version of which
was sculptured by students attending the Kevo Winter School to guard
Vortexland---stands for superfluid phase transitions, which can be
described either in the conventional (duck) way \cite{kevo}, or in the
Onsager (rabbit) way pursued here.

The next section discuses cluster percolation---the paradigm of
geometrical phase transitions---in its pure form known as
\textit{uncorrelated} percolation (Sec.\ \ref{sec:clusters}) and in its
applied form known as \textit{correlated} percolation to describe
thermal phase transitions in spin models (Sec.\ \ref{sec:corper}) and
the liquid-gas transition (\ref{sec:lg}).  Section \ref{sec:RW} gives a
parallel treatment of random walks, in its pure form
(\ref{sec:brownian}) and in its applied form to describe random
(\ref{sec:rvt}) and correlated vortex tangles featuring in superfluid
phase transitions (\ref{sec:cvt}) and also in superfluid turbulence and
defect formation after a rapid phase transition (\ref{sec:examples}).
The section is set up such as to highlight the close similarity between
cluster percolation and loop proliferation.  Section \ref{sec:va}
describes vortex lines in detail, starting with noninteracting vortices
(\ref{sec:schr}), followed by including interactions (\ref{sec:inter})
and an external field (\ref{sec:exfield}).  Section \ref{sec:entangled}
discusses entangled vortex lines as they appear when the Abrikosov flux
lattice melts (\ref{sec:melting}) from the perspective of Feynman's
cooperative exchange ring theory of Bose-Einstein condensation
(\ref{sec:rings}) and also discusses the order parameter describing this
phase (\ref{sec:op}).

\section{Clusters}
\label{sec:clusters}
In this section, the paradigm of geometrical phase transitions, viz.\
percolation, and its extension to describe thermal phase transitions are
discussed.
\subsection{Uncorrelated Percolation}
\label{sec:up}
\begin{figure}
\begin{center}
\includegraphics[width=4.0cm]{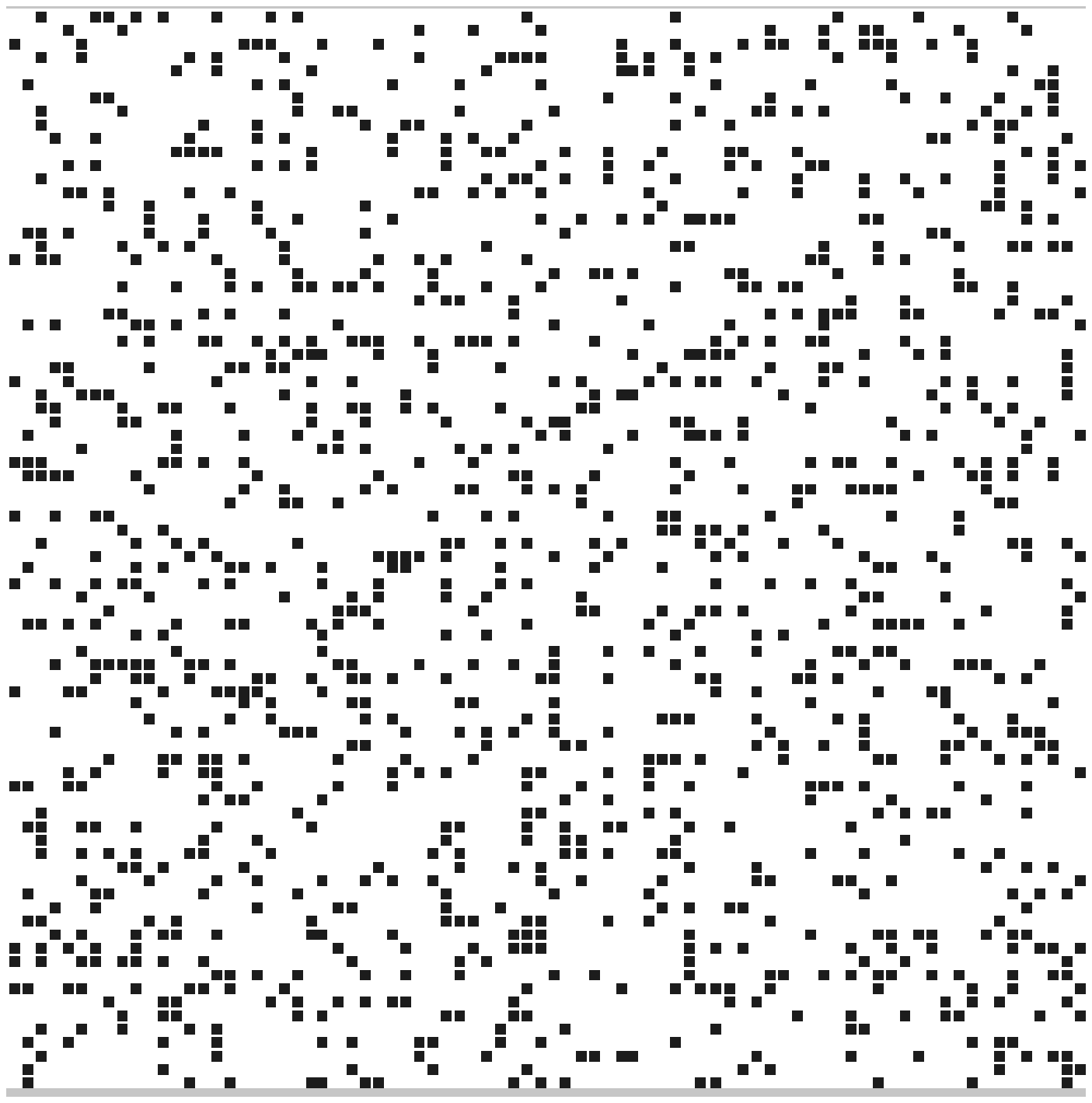} 
\hspace{2cm}
\includegraphics[width=4.0cm]{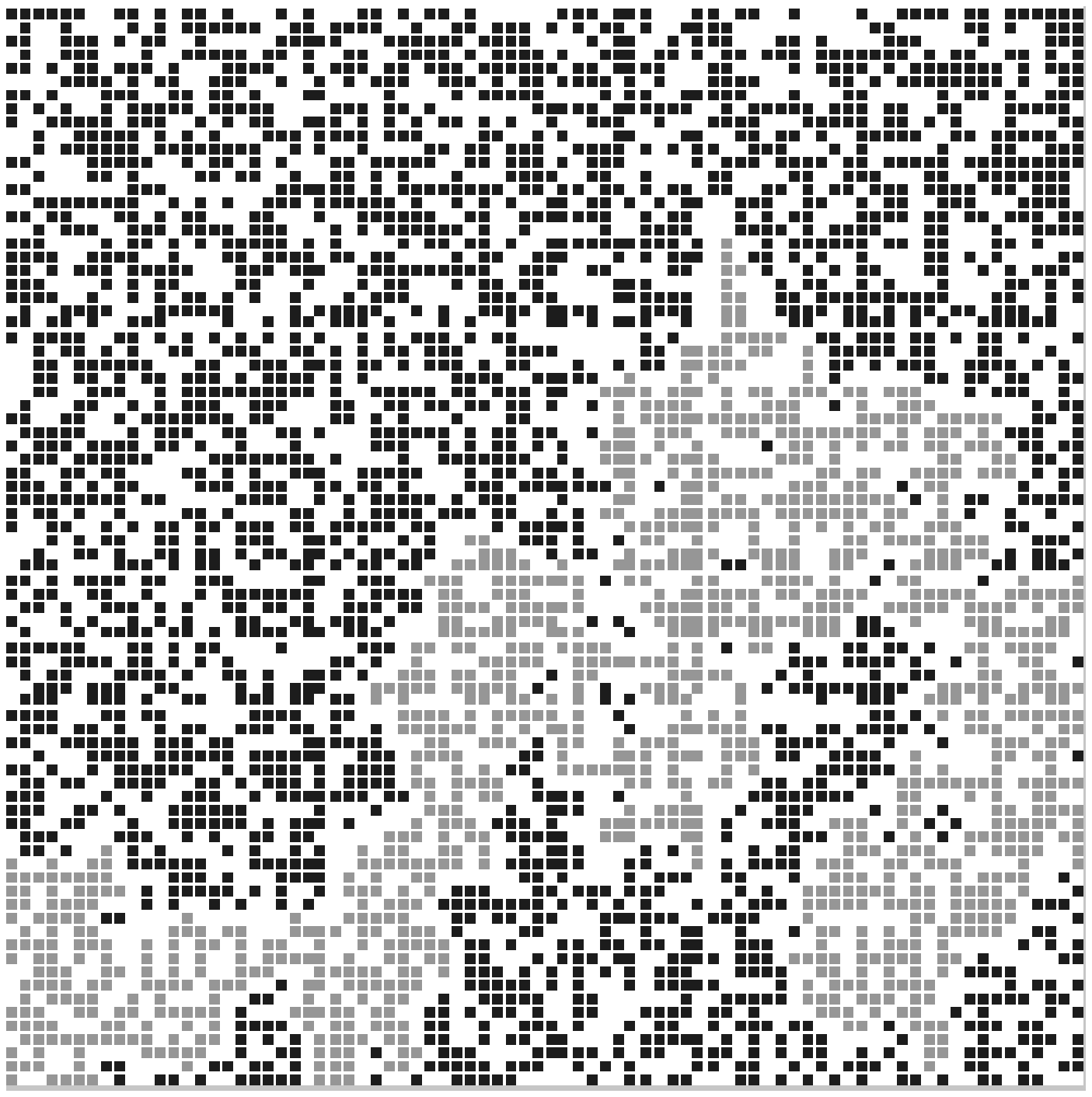} 
\end{center}
\caption{Typical output \cite{applet} of the site percolation algorithm
implemented on a two-dimensional square lattice for $p<p_{\rm c}$ (left
panel) and $p \sim p_{\rm c}$ (right panel).  Occupied sites are marked
by a dark square, with those belonging to the percolating cluster having
a lighter grayscale, while unoccupied sites are unmarked.
\label{fig:per_shot}}
\end{figure}

To define (site) percolation, consider a lattice with $N$ sites labeled
by the index $1 \leq i \leq N$.  Imagine visiting a site and generating
a random number $0<r_i<1$.  If that number is smaller than a predefined
number $p$, the occupation probability, occupy the site, otherwise leave
it unoccupied.  This algorithm can be summarized in pseudo computer code
as:
\texttt{
\begin{quote}
\begin{tabbing}
wh\=ile $i < N$ \\ 
\> generate $0<r_i<1$ \\ 
\> if $r_i < p$ occupy site \\ 
\> $i++$ \\ end while
\end{tabbing}
\end{quote}
}
\noindent 
After all lattice sites have been visited in this way, a random pattern
of occupied and unoccupied sites emerges (see Fig.~\ref{fig:per_shot}).
Next-neighboring occupied sites can be grouped together into clusters,
whose properties as a function of $p$ form the subject of percolation
theory.  Since a given site is occupied whether or not neighboring (or
any other) sites are, this process is more precisely referred to as {\it
uncorrelated} percolation.

As the occupation probability $p$ increases, more sites become occupied
and the clusters grow.  For high-enough values of $p$, a large cluster,
spanning the lattice is frequently generated during updates (see the
right panel of Fig.~\ref{fig:per_shot}).  The appearance of such a
percolating cluster is a fundamental property and not an artifact of
working on a finite lattice.  Even on an infinite lattice, a percolating
cluster, which would then be of infinite size, appears.  The only
difference between a finite and infinite lattice is that the percolating
cluster on an infinite lattice first appears at a specific value $p_{\rm
c}$, special to the dimensionality and symmetry of the lattice
considered, whereas on a finite lattice there is always a finite
probability to generate a percolating cluster when $p<p_{\rm c}$.  Also,
for $p> p_{\rm c}$, the percolating cluster is always generated on an
infinite lattice, but not necessarily so on a finite one.  The theory of
finite-size scaling is needed to extract the value of $p_{\rm c}$ from
numerical simulations done on finite lattices \cite{BinderHeermann}.

Percolation theory \cite{StauferAharony} can be set up starting from the
cluster size distribution $\ell_s (p)$, giving the number density of
clusters of size $s$.  Close to the percolation threshold, it takes the
form
\begin{equation} 
\label{perdis}
\ell_s (p) \propto s^{- \tau} \, \mathrm{e}^{- c s}, \quad \quad c \propto
(p_{\rm c} - p)^{1/\sigma},
\end{equation} 
where the coefficient $c$ vanishes with an exponent $1/\sigma$ when the
percolation threshold $p_{\rm c}$ is approached from below.  The cluster
size distribution (\ref{perdis}) contains two factors.  The first one
measures (as will become clear below when discussing random walks) the
configurational entropy of clusters, while the exponential is similar to
a Boltzmann factor which for finite $c$ suppresses large clusters.  The
latter vanishes when the percolation threshold is approached from below,
implying no longer a restriction on forming arbitrary large clusters and
resulting in a proliferation of clusters. 

Although percolation lacks a conventional description in terms of an
Hamiltonian, a partition function $Z$ can nevertheless be defined
through
\begin{equation} 
\label{perZ}
\ln(Z) \propto \mathbb{V} \sum_s \ell_s,
\end{equation} 
with $\mathbb{V}$ denoting the volume of the system.  The right hand counts
the number of clusters of all sizes.  As in statistical physics, various
physically relevant quantities can be calculated from $Z$, such as the
percolation strength $P(p)$, which for $p > p_{\rm c}$ denotes the
probability that a randomly chosen site belongs to the percolating
cluster, and the average cluster size $S(p)$.  Specifically,
\begin{equation} 
\label{PS}
P = \frac{\partial \ln(Z)}{\partial c}, \quad S = \frac{\partial^2
\ln(Z)}{\partial c^2},
\end{equation} 
showing that the percolation strength $P$ is similar to the
magnetization in spin models, while the average cluster size $S$ is
similar to the magnetic susceptibility.  Near the percolation threshold,
the quantities (\ref{PS}) show power-law behavior, $P(p) \sim (p- p_{\rm
c})^{\beta_{\rm per}}$ and $S(p) \sim |p_{\rm c} - p|^{- \gamma_{\rm
per}}$, characterized by the critical exponents $\beta_{\rm per}$ and
$\gamma_{\rm per}$, where the subscript ``per'' is to indicate that they
pertain to percolation.  A final critical exponent, $\eta_{\rm per}$,
specifies the power-law behavior of the correlation function $G_{\rm
per}$ at the percolation threshold.  Physically, $G_{\rm per}({\bf x},
{\bf x}')$ denotes the probability that the sites with position vectors
${\bf x}$ and ${\bf x}'$ belong to the same cluster.  It usually is only
a function of the distance between the two sites and has at $p=p_{\rm
c}$ the algebraic behavior in $d$ space dimensions
\begin{equation} 
\label{algebraic}
G_{\rm per}(x) \sim \frac{1}{x^{d-2 + \eta_{\rm per}}}.
\end{equation} 
The exponent $\sigma$, together with the so-called Fisher exponent
\cite{Fisher} $\tau$, specifying the cluster size distribution
(\ref{perdis}), determine these critical exponents through scaling
relations [see Eq.\ (\ref{pec}) below].

Since the percolating cluster, often also referred to as
\textit{infinite} cluster, includes a macroscopic fraction of the total
number of sites, its mere size would dominate any sum.  For this reason,
summations $\sum_s$ over cluster sizes exclude the percolating, or
infinite one.  In practice \cite{BinderHeermann}, when working on a
finite lattice, simply the largest cluster present is excluded, whether
or not is spans the lattice.  The percolation strength $P$ can
nevertheless be extracted from considering only finite clusters in Eq.\
(\ref{Zlg}) because of the constraint
\begin{equation}
\label{constraint} 
p = P(p) + \sum_s s \ell_s(p),
\end{equation} 
stating that an occupied site either belongs to the infinite cluster
or to a finite one.   This constraint is special to percolation.

Another important quantity characterizing percolation is the radius of
gyration $R_s$,
\begin{equation} 
R_s^2 = \frac{1}{s} \sum_{i=1}^s ({\bf x}_i - \bar{\bf x})^2,
\end{equation} 
which gives a measure of the spatial extent of a cluster of size $s$
centered at $\bar{\bf x} = (1/s) \sum_{i=1}^s {\bf x}_i$.  For large
enough clusters, $R_s$ scales with $s$ as
\begin{equation} 
\label{Hausdorff}
R_s \sim s^{1/D},
\end{equation} 
defining the Hausdorff, or fractal dimension $D$.  Close to the
percolation threshold, $D$ can be related to the correlation length
exponent $\nu$, specifying how the correlation length $\xi$ diverges,
$\xi(p) \sim |p_{\rm c} - p|^{- \nu}$.  The relation reads
\cite{StauferAharony}
\begin{equation}
\label{nu} 
\sigma = \frac{1}{D \nu}.
\end{equation} 
Also the Fisher exponent $\tau$ can be easily related to the fractal
dimension, giving 
\begin{equation}
\label{tau} 
\tau = \frac{d}{D} + 1,
\end{equation}  
where $d$ denotes the number of space dimensions.  

Using scaling laws, one can express all the critical exponents in terms
of just the two exponents $\sigma$ and $\tau$, specifying the cluster
size distribution (\ref{perdis}).  For example \cite{StauferAharony},
\begin{equation}
\label{pec} 
\beta_{\rm per} = \frac{\tau -2}{\sigma}, \quad \gamma_{\rm per} =
\frac{3-\tau}{\sigma}, \quad   \eta_{\rm per} = 2 + d - 2 D,
\end{equation} 
with $\sigma$ increasing from $36/91 \approx 0.40$ on a two-dimensional
square lattice to $1/2$ on a six-dimensional square lattice, while at the
same time, the fractal dimension increases from $D = 91/48 \approx 1.90$
to $4$.  Six is the upper critical dimension of uncorrelated
percolation, beyond which the exponents remain fixed at their mean-field
values $\sigma=1/2$ and $D=4$ \cite{StauferAharony}.

\begin{figure}[b]
\begin{center}
\includegraphics[width=6.0cm]{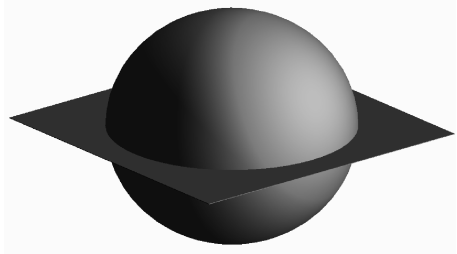} 
\end{center}
\caption{A 2-sphere denoting the possible values a spin of the O(3),
or Heisenberg model can take.  A randomly chosen vector ${\bf r}$ (not
shown in the figure) can be used to cut the sphere in two halves,
depending on the sign of the projections of a spin onto that vector.
\label{fig:cut_ball}}
\end{figure}

The discussion so far concentrated on clusters of occupied sites, which
for $p<p_{\rm c}$ make up the minority phase, while the unoccupied sites
form the majority phase.  At the percolation threshold this changes and
the roles of the minority and majority phases are interchanged.  In
other words, when approaching the percolation threshold from above, one
would naturally concentrate on the disfavored unoccupied, rather than on
the abundant occupied sites.  From that side, the transition corresponds
to a proliferation of clusters of unoccupied sites.  In a sense, the two
descriptions are dual to each other with the minority and majority
phases interchanged.

\subsection{Correlated Percolation}
\label{sec:corper}
To see how percolation theory can be applied to thermal phase
transitions taking place in spin models consider the Ising model at the
absolute zero of temperature \cite{StauferAharony}.  Here, in the
absence of fluctuations, all the spins, which in principle can take the
values $S_i = \pm 1$, point in one direction.  When the temperature $T$
increases, some of the spins will reverse their direction.  The
resulting spin configuration at a given temperature is easily mapped
onto a percolation problem by identifying the sites with reversed spins
with occupied sites.  The question is now if the critical behavior of
the Ising model can be recovered, focusing exclusively on these clusters
instead of the magnetic properties.  In other words, are the correct
values for the critical temperature and the critical exponents obtained
when measuring quantities like the percolation strength and the average
cluster size discussed above?

It turns out that, in general, clusters of sites with reversed spins
proliferate already before the thermal phase transition is reached.
(Only on a two-dimensional square lattice, the temperature where a
percolating cluster first appears coincides with the critical
temperature.)  Moreover, the critical exponents thus obtained are those
of uncorrelated percolation and not of the Ising model (even on a
two-dimensional square lattice).  This means that the clusters constructed
in this naive way lack essential information about the Ising model and
need to be modified.

To construct a modified cluster \cite{CK}, take two nearest-neighboring
sites $i, j$ of a naive cluster with its spins reversed, and include the
pair into the modified cluster only with the bond probability
\begin{equation} 
p_{i j} = 1 - \exp(-2 \beta J),
\end{equation} 
where $J$ is the spin-spin coupling of the Ising model and $\beta =
1/k_{\rm B} T$, with $k_{\rm B}$ Boltzmann's constant.  The factor $2J$
appearing here corresponds to the increase in energy when one of the two
spins involved in the bond is flipped.  The upshot of this modification
is that the resulting bond clusters are in general smaller than the
naive ones and also more loosely connected.  The modified clusters turn
out to proliferate right at the critical temperature and display
critical behavior identical to that of the Ising model---as should be
according to an exact map of the partition function of the Ising model
onto correlated percolation by Fortuin and Kasteleyn
\cite{FortuinKasteleyn}.

The above scenario was recently extended to three-dimensional O($n$) spin
models, with $n=2,3,4$.  Instead of taking only the two values $S_i =
\pm 1$, which can be thought of as comprising a zero-dimensional sphere,
the spins of the O($n$) model can take any value on the
($n$$-$1)-dimensional sphere defined by ${\bf S}_i^2 = 1$.  The O($n$)
model with $n=2$ corresponds to the XY-model, while that with $n=3$
corresponds to the Heisenberg model.  An exact analytic map onto
correlated percolation as existed for the Ising model is not available
for $n>1$, so that numerical simulations are needed to establish the
existence of such a map \cite{Satzetal}.  Two nearest-neighboring sites
$i$ and $j$ with spin ${\bf S}_i$ and ${\bf S}_j$ are included in a
cluster with bond probability
\begin{equation} 
p_{i j}= 1- \mathrm{e}^{-2 \beta J \max[0, ({\bf S}_i \cdot {\bf r}) ({\bf
S}_j \cdot {\bf r})]},
\end{equation} 
where ${\bf r}$ is a randomly chosen unit vector which can be used to
cut the unit hypersphere ${\bf S}_i^2=1$ in two halves (see
Fig.~\ref{fig:cut_ball}).  When the spins ${\bf S}_i$ and ${\bf S}_j$
take values in opposite halves of the hypersphere, their projections
onto ${\bf r}$ have opposite sign.  Such a pair is never included in a
cluster since $p_{ij}=0$ then.  When the spins take values in the same
halve, the pair is included in a cluster with the bond probability
determined by the projection ${\bf S}_i \cdot {\bf r}$ of their spin
onto the random vector.  The clusters thus constructed turn out to
proliferate at the critical temperature where the thermal phase
transition takes place.  Moreover, the critical exponents obtained by
measuring the percolation strength and average size of these clusters
are the same as those obtained in the more traditional manner by
focusing on the magnetic properties of these spin models
\cite{Satzetal}.

These examples illustrate that thermal phase transitions taking place in
a wide range of statistical physics models can be reformulated in terms
of percolation theory.  Each of these models is thus characterized by a
partition function of the form (\ref{perZ}) with a cluster size
distribution $\ell_s$ given by Eq.\ (\ref{perdis}).  The exponents
$\sigma$ and $\tau$ characterizing $\ell_s$ take values specific for the
universality class to which the model belong.  Given the values of
$\sigma$ and $\tau$, which in general differ from those of uncorrelated
percolation, the critical exponents follow from scaling relations, as
in Eqs.\ (\ref{nu})-(\ref{pec}).

\subsection{Liquid-Gas Transition}
\label{sec:lg}
Next, the condensation of a classical gas, the liquid-gas transition, is
considered from the perspective put forward by Mayer \textit{et al.}
\cite{Mayer,MM} and shown to be closely related to correlated
percolation.

\begin{figure}
\begin{center}
\psfrag{p}[t][t][1][0]{$P$} 
\psfrag{t}[t][t][1][0]{$T$}
\psfrag{tc}[t][t][1][0]{$T_{\rm c}$}
\psfrag{l}[t][t][1][0]{liquid}
\psfrag{g}[t][t][1][0]{gas}
\includegraphics[width=6.0cm]{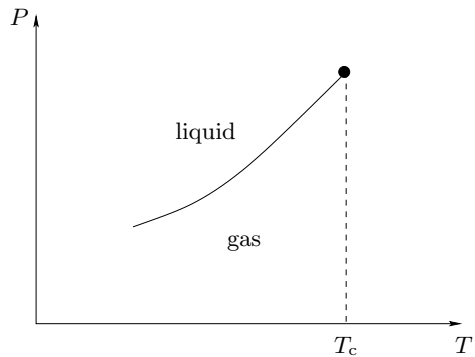} 
\end{center}
\caption{Sketch of part of the liquid-gas phase diagram, showing the
coexistence curve and the critical point at which this curve terminates.
\label{fig:pd}}
\end{figure}  
The relevant part of a typical phase diagram is sketched in
Fig.~\ref{fig:pd}.  Below a critical temperature $T_{\rm c}$, the gas
phase is separated from the liquid phase by a discontinuous phase
transition.  At the phase boundary, the two phases have the same free
energy and despite different particle number densities coexist,
separated by an interface.  The energy of the mixed state is higher than
that of the pure states because the formation of the surface requires
energy---the surface tension.  The coexistence curve ends at a critical
point where the surface tension vanishes and the transition becomes
continuous with Ising model critical exponents.  The order parameter,
distinguishing the liquid from the gas phase, is the difference in
particle number density.

In Mayer's theory, which was the first description of a phase transition
beyond mean-field theory \cite{interview}, the full interaction between
molecules is approximated by a pair interaction.  That is,
three-particle and higher interactions, which are irrelevant when
describing critical behavior, are ignored. The grand-canonical partition
function $Z(\beta,\alpha)$, with $\alpha = - \beta \mu$, can then be
written as an Ursell-Mayer cluster series \cite{Ursell,Mayer}
\begin{equation}
\label{Zlg} 
\ln(Z) = \mathbb{V} \sum_l \frac{b_l}{\lambda_\beta^{3l}} \, \mathrm{e}^{-
\alpha l},
\end{equation} 
where $\lambda_\beta =\hbar \sqrt{2 \pi \beta/ m}$, with $m$ the mass of
a molecule, is the de Broglie thermal wavelength, and the $b_l$'s, with
$b_1=1$, denote the Mayer cluster integrals \cite{MM}.  Being determined
by the pair potential, these integrals can in principle be calculated,
but, apart from small clusters, are in practice too complicated to
evaluate for realistic choices of the potential.  A (classical) ideal
gas corresponds to setting $b_l=0$ for $l>1$.  With the abbreviation
\begin{equation}
\label{abbr}  
\ell_l = \frac{b_l}{\lambda_\beta^{3l}} \, \mathrm{e}^{-\alpha l},
\end{equation} 
the partition function (\ref{Zlg}) can be written in a form similar to
Eq.\ (\ref{perZ}) for percolation, 
\begin{equation}
\label{Zlg'}  
\ln(Z) = \mathbb{V} \sum_l \ell_l,
\end{equation} 
while the expression for the number $N$ of molecules, at least in the
gas phase, becomes
\begin{equation} 
\label{Nlg}
N = - \frac{\partial}{\partial \alpha} \ln(Z) = \mathbb{V} \sum_l l \, \ell_l.
\end{equation} 
It is tempting to interpret $\ell_l$ as the number density of clusters
containing $l$ molecules.  However, the Mayer cluster integrals $b_l$
and therefore $\ell_l$ can be negative because the intermolecular
interaction $u(r)$ is attractive for large separation $r$, behaving
asymptotically as $u(r) \sim - 1/r^6$.  In the gas phase, where large
clusters are exponentially suppressed so that only small clusters exist,
the above interpretation is not obstructed too much \cite{Becker}.

As for percolation, a possible infinite cluster, containing a
macroscopic fraction of the total number $N$ of molecules present, is
excluded in summations $\sum_l$ over cluster sizes.  (In numerical
simulations, simply the largest cluster is best excluded, whether or not
it spans the lattice.)

Condensation, achieved, say, by reducing the volume at fixed $N$,
corresponds in this description to the formation of an infinite cluster.
The sum $\sum_l l \, \ell_l$ in Eq.\ (\ref{Nlg}) denotes the number
density of molecules in the gas phase, which is represented by the
finite clusters.  When the coexistence curve is reached at the volume
$\mathbb{V}_{\rm co^+}$, the sum saturates at the value $N/\mathbb{V}_{\rm
co^+}$.  Upon reducing the volume further, an infinite cluster, which is
not accounted for in the summation $\sum_l$ over cluster sizes, must
form to accommodate the difference between the true density $N/\mathbb{V}$
and that contained in the gas phase, $N/\mathbb{V}_{\rm co^+}$.  For this to
be possible, the exponential suppression of large clusters needs to
vanish.  Since the chemical potential remains negative at the
coexistence curve, the Mayer cluster integrals must yield an
$l$-dependent exponential factor to cancel the factor $\exp(-\alpha l)$
in Eq.\ (\ref{abbr}) at that curve.  That is,
\begin{equation} 
\frac{b_l}{\lambda_\beta^{3l}} \propto l^{-\tau} \, \mathrm{e}^{\alpha_{\rm
co} l}, 
\end{equation} 
close to the coexistence curve, with the first factor measuring the
configurational entropy of clusters \cite{f1}.  It then follows that
\begin{equation} 
\label{dislg}
\ell_l \propto l^{-\tau} \, \mathrm{e}^{-(\alpha - \alpha_{\rm co})l},
\end{equation} 
so that as $\alpha$ approaches $\alpha_{\rm co}$ from above, $\ell_l$
displays algebraic behavior, as required.  This limit also marks the
radius of convergence of the sum (\ref{Nlg}): for $\alpha$ smaller than
$\alpha_{\rm co}$, the sum would diverge, i.e., at the coexistence
curve \cite{Becker}
\begin{equation} 
\label{radius}
\lambda_\beta^3 \, \mathrm{e}^\alpha = \lim_{l \to \infty} (l b_l)^{1/l}.
\end{equation} 

The discontinuous liquid-gas transition fundamentally differs from the
continuous phase transitions discussed so far.  In the earlier
discussion, the appearance of an infinite cluster signaled the
(continuous) transition of the entire system from one phase into the
other.  That is, the infinite cluster \textit{together} with the smaller
clusters still present (see the right panel of Fig.~\ref{fig:per_shot})
represent the new phase.  In the liquid-gas transition, the situation
differs in that the infinite cluster denotes the liquid phase, while the
smaller clusters denote the gas phase.  In other words, the right panel
of Fig.~\ref{fig:per_shot} now represents a snapshot of a
two-dimensional section of the system (in microgravity conditions, say)
with both the liquid phase (percolating cluster) and the gas phase
(smaller clusters) coexisting.

This interpretation is supported by the observation that the pressure
associated with the finite clusters satisfies the Clausius-Clapeyron
equation \cite{Becker}.  To first order, the pressure $P= \ln(Z)/\beta
\mathbb{V}$ in the gas phase can be approximated by that of an ideal gas,
corresponding to the first term in the Ursell-Mayer expansion
(\ref{Zlg}).  When differentiated with respect to $T$, this gives
\begin{equation} 
\label{dP}
\frac{\mathrm{d} \ln (P)}{\mathrm{d} T} = \frac{1}{T} - \lim_{l \to \infty}
\frac{1}{l} \frac{\mathrm{d} \ln (b_l)}{\mathrm{d} T} ,
\end{equation} 
where use is made of Eq.\ (\ref{radius}).  To interpret the last term,
the energy $E= - \mathrm{d} \ln (Z)/\mathrm{d} \beta$ of the gas phase
is determined, where $\alpha$ is to be considered a $\beta$-independent
variable, leading to the result
\begin{equation} 
\label{Edif}
E = \mathbb{V} \sum_l \epsilon_l l \ell_l ,
\end{equation} 
with $\epsilon_l$ the energy per molecule of a cluster of size $l$,
\begin{equation} 
\label{epsilon}
\epsilon_l = \tfrac{3}{2} k_{\rm B} T + k_{\rm B} T^2 \frac{1}{l}
\frac{\mathrm{d} \ln (b_l)}{\mathrm{d} T}.
\end{equation} 
Since $\mathrm{d} b_l/\mathrm{d} T$ is in general negative, $\epsilon_l$
is smaller than the average kinetic energy per molecule, $\frac{3}{2}
k_{\rm B} T$.  Physically, the last term in Eq.\ (\ref{epsilon}) denotes
minus the (average) binding energy $\upsilon_l$ per molecule.  In terms
of this quantity, Eq.\ (\ref{dP}) becomes the Clausius-Clapeyron
equation \cite{Becker}
\begin{equation} 
\frac{\mathrm{d} \ln (P)}{\mathrm{d} T} = \frac{k_{\rm B}T +
\upsilon_\infty}{k_{\rm B} T^2},
\end{equation} 
with the numerator at the right hand denoting the energy required to
vaporize one molecule from the liquid, viz.\ the sum of the thermal
energy $k_{\rm B} T$ and binding energy $\upsilon_\infty$.

It should be kept in mind that the liquid-gas transition, being
discontinuous, is special in that both the liquid and gas phase can
coexist, each occupying separate parts of the total volume.  Similar
arguments as the ones developed for the liquid-gas transition are
sometimes applied to an ideal Bose gas to argue that Bose-Einstein
condensation is a discontinuous phase transition.  A Bose-Einstein
condensate is, however, formed in the zero-momentum state invading all
of space and therefore intertwined with the non-condensed part.  No
interface separating the two parts exist.  The situation is similar to
that in the O($n$) spin models discussed above, where, at the critical
temperature, the entire system transforms into a new phase that cannot
be spatially resolved into two separates parts, each occupying its own
volume.

Recently, the three-dimensional three-state Potts model in an an
external field has been numerically investigated, using clusters as
fundamental objects \cite{FortunatoSatz}.  The phase diagram of this
model is very similar to that of the liquid-gas transition, having a
line of discontinuous phase transitions at small fields that terminates
in a point at some critical field value where the transition becomes
continuous with Ising model exponents.

\section{Random walks}
\label{sec:RW}
The discussion of phase transitions so far was modeled after percolation
theory, with clusters specified by the cluster size distribution
(\ref{perdis}) playing the central role.  Another paradigm of thermal
phase transitions is random walks to which the discussion now turns.
\subsection{Brownian Random Walks}
\label{sec:brownian}
The simplest random walk is the Brownian random walk
\cite{KleinertPath}.  On a $d$-dimensional square lattice, where each
site has $2d$ nearest neighbors, a Brownian random walker located at a
given site $i$ randomly chooses among its nearest neighbors (n.n.) $m_i
\in \{\pm {\bf j}\}$ the next site to visit.  Here, the $d$ unit vectors
spanning the square lattice are denoted by ${\bf j}$.  The random walker
is not subject to any restriction and can, for example, return to a site
it visited before.  In pseudocode, the algorithm describing a Brownian
random walk reads: \texttt{
\begin{quote}
\begin{tabbing}
wh\=ile $i < n$ \\
\> generate $m_i \in \{\pm {\bf j}\}$ \\
\> move to n.n.\ $m_i$ \\
\> $i++$ \\ 
end while 
\end{tabbing}
\end{quote}
}
\begin{figure}
\begin{center}
\psfrag{r}[t][t][1][0]{$R_n$} 
\includegraphics[width=5.0cm]{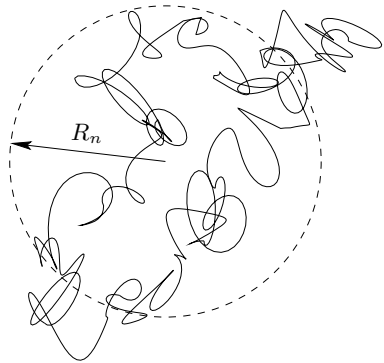} 
\end{center}
\caption{Schematic representation of the radius of gyration $R_n$ of a
closed random walk containing $n$ steps.
\label{fig:rs_rw}}
\end{figure}

\noindent 
A central concept in the theory of random walks is the probability
$K_n({\bf x} \to {\bf x}')$ for a random walker to move from one site
with position vector ${\bf x}$ to another ${\bf x}'$ in $n$ steps.  The
probability is given by
\begin{equation} 
\label{Kn}
K_n({\bf x} \to {\bf x}') = \frac{\#\mbox{ of paths } {\bf x} \to {\bf
x}' \mbox{ in } n \mbox{ steps}}{\#\mbox{ of paths } {\bf x} \to \; \star \;
\mbox{ in } n \mbox{ steps}},
\end{equation} 
where the denominator, counting paths with arbitrary endpoints
(indicated by $\star$), is included so that $K_n$ is a genuine
probability, taking values between 0 and 1.  To evaluate it, imagine the
random walker taking a first step to one of its nearest neighbors $\pm
{\bf j}$.  It then has only $n-$1 steps still available to reach the
final destination ${\bf x}'$, with probability $K_{n-1}({\bf x}+ a {\bf
j} \to {\bf x}')$.  That is, $K_n$ satisfies the recurrence relation:
\begin{equation} 
\label{recu}
K_n({\bf x} \to {\bf x}') = \frac{1}{2d} \sum_{\pm {\bf j}} K_{n-1}({\bf
x}+ a {\bf j} \to {\bf x}'),
\end{equation} 
where the factor $2d$ denotes the number of nearest neighbors on a
$d$-dimensional square lattice.  When $K_{n-1}({\bf x} \to {\bf x}')$ is
subtracted from both sides of Eq.\ (\ref{recu}), a difference equation
is obtained
\begin{equation} 
K_n({\bf x} \to {\bf x}') - K_{n-1}({\bf x} \to {\bf x}') = \frac{1}{2d}
\sum_{\pm {\bf j}} \left[K_{n-1}({\bf x}+ a {\bf j} \to {\bf x}') -
K_{n-1}({\bf x} \to {\bf x}') \right],
\end{equation} 
with the obvious boundary condition $K_0 ({\bf x}\to {\bf x}') =
\delta_{{\bf x},{\bf x}'}$.  The continuum limit is taken by
simultaneously letting $a \to 0$ and $n \to \infty$, such that the
length $L = na$ of the random walk satisfies $L a = na^2 \to
\mbox{const}$.  The difference equation then turns into the differential
equation
\begin{equation} 
\label{diff}
\partial_n K_n({\bf x} \to {\bf x}') = \frac{a^2}{2d} \nabla^2 K_n({\bf
x} \to {\bf x}'),
\end{equation} 
with the celebrated solution
\begin{equation} 
\label{celebrated}
K_n({\bf x} \to {\bf x}') = \left( \frac{d}{2 \pi n a^2} \right)^{d/2}
\exp \left[- \frac{d}{2} \frac{({\bf x} - {\bf x}')^2}{n a^2}\right],
\end{equation} 
where $na^2 \to \mbox{const}$.  Each dimension the random walker is free
to roam gives a contribution $\sqrt{d/2 \pi n a^2}$ to the prefactor.

The radius of gyration $R_n$ of a random walk containing $n$ steps is
defined by 
\begin{equation} 
\label{Rn}
R_n^2 = \frac{1}{n} \sum_{i=1}^n ({\bf x}_i - \bar{\bf x})^2.
\end{equation} 
As for clusters, it measures the spatial extent of the random walk
containing $n$ steps and centered at $\bar{\bf x} = (1/n) \sum_{i=1}^n
{\bf x}_i$ (see Fig.~\ref{fig:rs_rw}).

It scales with the number of steps as
\begin{equation} 
\label{fractal}
R_n \sim n^{1/D},
\end{equation} 
where for Brownian random walks, the fractal dimension is $D=2$.

When the random walk is taken as representing a flexible polymer or
vortex, the energy to create such a line object needs to be included.
Let $\theta$ denote that energy per unit length, i.e., the line tension,
then this is achieved by the modification
\begin{equation} 
\label{modification}
K_n({\bf x} \to {\bf x}') \to K_n({\bf x} \to {\bf x}') \, \mathrm{e}^{-
\beta \theta n a}.
\end{equation} 
Because of the energy cost, the Boltzmann factor exponentially
suppresses long lines, provided the line tension is positive.  The total
probability for a random walker to move from one site ${\bf x}$ to
another ${\bf x}'$ using an arbitrary number of steps defines the
correlation function $G({\bf x}, {\bf x}')$, which is assumed to depend
only on the distance between the two sites:
\begin{equation}
\label{G} 
G(|{\bf x} - {\bf x}'|) = \sum_n K_n({\bf x} \to {\bf x}') \, {\rm
e}^{- \beta \theta n a}.
\end{equation} 

In the absence of external sources and boundaries, vortex lines cannot
terminate inside the system and must form closed loops.  The partition
function describing a (grand canonical) ensemble of closed,
noninteracting vortex lines is given by
\begin{equation} 
\ln (Z) = \mathbb{V} \sum_n \frac{1}{n} K_n({\bf x} \to {\bf x}) \,
\mathrm{e}^{- \beta \theta n a},
\end{equation} 
where the notation $K_n({\bf x} \to {\bf x})$ indicates that the random
walk representing the vortex line starts and ends at the same point.
The right hand of the equation is therefore ${\bf x}$-independent, as
required for a partition function.  Since a closed loop of length $na$
can be traversed starting from any of the $n$ sites visited, the factor
$1/n$ prevents overcounting.  With the explicit solution
(\ref{celebrated}), the partition function becomes
\begin{equation}  
\ln (Z) = \mathbb{V} \sum_n \ell_n,
\end{equation} 
where 
\begin{equation} 
\label{loopdis}
\ell_n \propto n^{-\tau} \, \mathrm{e}^{- \beta \theta n a}, \hspace{2ex}
\tau = \frac{d}{D} + 1.
\end{equation} 
Physically, $\ell_n$ denotes the loop size distribution, giving the
number density of loops containing $n$ steps.  In this way, the
partition function is represented in a form surprisingly similar to that
in percolation theory Eq.\ (\ref{perZ}) and for the liquid-gas
transition Eq.\ (\ref{Zlg'}), with the proviso that not clusters, but
loops are considered.
\subsection{Random Vortex Tangles}
\label{sec:rvt}
Given the close similarity in describing uncorrelated clusters and
random vortex loops, the question arises whether both systems are in the
same universality class.  To settle this question numerically, an
algorithm similar to that for uncorrelated percolation is needed to
generate a random vortex tangle, consisting of intertwined closed vortex
loops.  Such an algorithm is provided by the Vachaspati-Vilenkin
algorithm \cite{VachaspatiVilenkin} developed to study defect formation
in rapid (cosmic) phase transitions \cite{leshouches}.

\begin{figure}[b]
\begin{center}
\psfrag{0}[t][t][1][0]{0} 
\psfrag{2}[t][t][1][0]{2} 
\psfrag{4}[t][t][1][0]{4} 
\psfrag{v}[t][t][1][0]{vortex} 
\psfrag{a}[t][t][1][0]{a} 

\includegraphics[width=8.0cm]{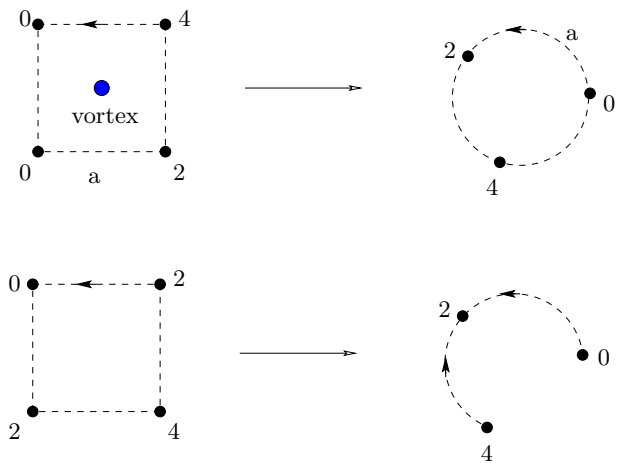} 
\end{center}
\caption{Illustration of Kibble's geodesic rule
\cite{Kibble}. \textit{Top}: Plaquette pierced by a vortex.  The path
segments labeled by ``a'' show the path taken by the phase along the
vacuum manifold (circle at the right) when one moves from the left-lower
to the right-lower corner on the plaquette circling the vortex.
\textit{Bottom}: Although all three available values of the phase
appear, no vortex pierces the plaquette.  When the right-lower corner is
reached, where the phase is $4\pi/3$, the phase moves back to $2\pi/3$
rather than to $0$ to complete the circle, representing the vacuum
manifold.
\label{fig:exs}}
\end{figure}

Quantized vortex lines emerge when a U(1) symmetry is spontaneously
broken \cite{PWA}.  The vacuum of a state with such a broken symmetry
is characterized by a phase factor $\exp(i\varphi)$, with the phase
$\varphi$ changing by $2\pi$ when a vortex of unit topological charge is
circled once.

Consider discretizing both space (to a three-dimensional square lattice,
say, with $N$ sites) and also the vacuum manifold, so that $\varphi$
takes only certain discrete values in the interval $0 \leq \varphi < 2
\pi$.  The restriction to this interval assures that only vortices of
unit strength are generated.  Imagine visiting a site of the lattice and
randomly assigning it one of the discrete phase values.  The random
choice guarantees that phases at different lattice sites are
uncorrelated.  The Vachaspati-Vilenkin algorithm, with $\varphi$ taking
the minimum number three of different values, can be summarized in
pseudocode as:
\texttt{
\begin{quote}
\begin{tabbing}
wh\=ile $i < N$ \\ \> generate $\varphi_i \in \left\{0, 
\frac{2}{3} \pi, \frac{4}{3} \pi \right\}$ \\ \> $i++$ \\
end while
\end{tabbing}
\end{quote}
}
\noindent 
After all lattice sites have been visited in this way, vortices are
traced by going around each plaquette of the lattice once.  When upon
returning to the starting site, a phase difference of $2\pi$ is found,
it is concluded that a vortex penetrates the plaquette.  In going from
one lattice site with the phase $\varphi_1$ to a neighboring site with
the phase $\varphi_2$, Kibble's geodesic rule \cite{Kibble} is used to
select the shortest path along the circle representing the vacuum
manifold connecting $\varphi_1$ and $\varphi_2$ (see Fig.~\ref{fig:exs}
for two examples \cite{VachaspatiVilenkin}).

\begin{figure}
\begin{center}
\includegraphics[width=6.0cm]{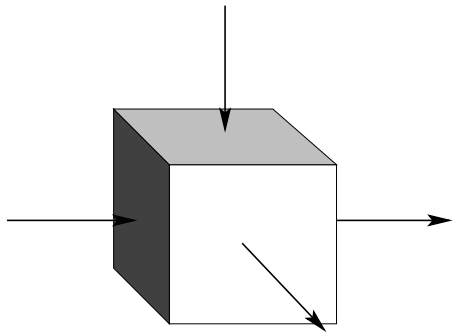} 
\end{center}
\caption{Illustration of two vortices entering a (three-dimensional)
unit cell.
\label{fig:assign}}
\end{figure}

When two vortices are found to enter a unit cell (see
Fig.~\ref{fig:assign}), the two incoming and two outgoing vortex
segments are randomly connected.
After all the plaquettes of the lattice have been considered, a random
vortex tangle, consisting of closed loops can be traced out.  

The fractal dimension of the vortex tangle found in the initial
simulations \cite{VachaspatiVilenkin} were consistent with that of
Brownian random walks, $D=2$.  This result can be understood
\cite{HindmarshStrobl} by realizing that a dense random vortex tangle
was simulated, where a single vortex experiences an effective repulsion
from the neighboring vortex segments as the volume occupied by these
segments is no longer available.  This is similar to a polymer in a
dense solution, which also executes a Brownian random walk.

In the Vachaspati-Vilenkin algorithm, the allowed phase values have
equal probability so that the vortex tangle generated has always the
same average vortex line density.  To be able to vary the density,
Vachaspati \cite{Vachaspati} lifted the degeneracy of the vacuum
manifold by giving one phase value a bias, $p_{2\pi/3} = p_{4\pi/3} \neq
p_0$.  This can be achieved by choosing for example
\cite{HindmarshStrobl}
\begin{equation} 
p_{2\pi/3} = p_{4\pi/3} = C \mathrm{e}^{\eta/2}, \; p_0 = C
\mathrm{e}^{-\eta},
\end{equation} 
with $C^{-1} = 2 \exp(\eta/2)+ \exp(-\eta)$.  Since all phase values are
needed for a vortex to pierce a given plaquette, the bias obstructs the
formation of vortices and decreases the vortex line density.  The bias
parameter $\eta$ plays a role similar to that of the occupation
probability $p$, determining the cluster density in percolation theory.

Imagine starting in the regime with only small vortex loops present and
gradually changing the bias so that more and at the same time larger
loops appear.  It turns out that for small enough values of $\eta$ an
``infinite'' vortex loop is frequently generated during updates
\cite{Vachaspati}.  The division between finite and infinite vortex
loops is, of course, somewhat arbitrary on a finite lattice.  A possible
choice is to classify vortices with a length larger than \cite{ACR}
\begin{equation} 
\label{lc}
L_\mathrm{c} = \frac{d}{2 \pi} \frac{L^2}{a}
\end{equation} 
as infinite, where $L$ is the lattice size.  The length scale
$L_\mathrm{c}$ is the average length of Brownian random walks traversing
the lattice once.  In other words, most infinite loops wrap themselves
around the lattice and thus have a nonzero winding number.  Universal
quantities such as the critical exponents should be independent of the
precise choice.  On an infinite lattice, a threshold $\eta=\eta_{\rm c}$
exists, above which infinite vortex loops are absent and large, but
finite loops exponentially suppressed.  The sudden appearance of an
infinite vortex loop as the threshold $\eta=\eta_{\rm c}$ is approached
from above is similar to that of a percolating cluster in percolation
theory \cite{Vachaspati}.  For an infinite vortex to appear, the line
tension $\theta$ needs to vanish, so that the Boltzmann factor in the
loop size distribution (\ref{loopdis}) becomes unity and no longer
suppresses large loops, i.e.,
\begin{equation} 
\ell_n (\eta_{\rm c}) \propto n^{- \tau}.
\end{equation} 
In analogy with percolation theory, cf.\ Eq.\ (\ref{perdis}), the
vanishing of the line tension is specified by an exponent $\sigma$,
\begin{equation} 
\theta \propto (\eta - \eta_{\rm c})^{1/\sigma}.
\end{equation} 
At the threshold, vortex loops proliferate in the same way that clusters
do at the percolation threshold.

To characterize the transition, the same definition of critical
exponents can be used as in percolation theory, leading to the same
scaling relations (\ref{pec}) found there.  Numerical simulations of the
transition that random vortex tangles undergo when changing the bias
using this algorithm showed that the values for the critical exponents
are more or less consistent with those of three-dimensional uncorrelated
site percolation \cite{StroblHindmarsh}:
\begin{center}
\begin{minipage}{7cm}
\begin{ruledtabular}
\begin{tabular}{ccc}
critical & percolation & minimally \\
exponent & theory & discretized U(1) \\
\hline
$\sigma$ & 0.45 & 0.46(2) \\
$\beta_{\rm per}$ & 0.41 & 0.54(10) \\
$\gamma_{\rm per}$ & 1.80 & 1.59(10) 
\end{tabular}
\end{ruledtabular}
\end{minipage}
\end{center}
Other values obtained in simulations with the discretized vacuum
manifold given more than the minimal three points
\cite{StroblHindmarsh}, show even better agreement with percolation
theory, implying that uncorrelated percolation and random vortex tangles
are indeed in the same universality class.  In other words, the close
similarity in describing both phenomena discussed above extends to the
level where fluctuations, determining the numerical values of the
critical exponents, are included.

\subsection{Correlated Vortex Tangles}
\label{sec:cvt}
In Sec.\ \ref{sec:corper}, the extension of the theory of uncorrelated
percolation to describe thermal phase transitions in correlated systems
was discussed.  It was concluded that correlated and uncorrelated
percolation differ only in the specific values of the exponents $\sigma$
and $\tau$, specifying the cluster size distribution (\ref{perdis}).  The
general framework applies to both cases, with the various expressions
giving the critical exponents in terms of $\sigma$ and $\tau$ remaining
unchanged as they are protected by scaling laws.

Here, a similar extension is discussed for the random vortex tangle to
describe superfluid phase transitions.  An arbitrary---not just
random---vortex tangle is again described by the loop size distribution
(\ref{loopdis}) \cite{Copelandetal}.  In the superfluid phase, large
vortex loops are exponentially suppressed.  When the temperature
increases, larger loops appear and the vortex tangle becomes denser.  As
the critical temperature is approached, the line tension $\theta$
vanishes, thereby allowing vortex loops to proliferate and thus to
disorder the superfluid phase.  The vanishing of the line tension is
specified by the exponent $\sigma$,
\begin{equation} 
\theta \propto (T_{\rm c}-T)^{1/\sigma}.
\end{equation} 
As in correlated percolation, $\sigma$ can be related to the fractal
dimension $D$, introduced in (\ref{Rn}), leading to the same relation
(\ref{nu}).  The value of the fractal dimension differs in general from
the value obtained for a random vortex tangle or for Brownian random
walks ($D=2$).  At criticality, the correlation function (\ref{G}) shows
algebraic behavior,
\begin{equation} 
G(x) \sim \frac{1}{x^{d-2 + \eta}},
\end{equation} 
as does the correlation function (\ref{algebraic}) defined in
percolation theory.  Using scaling laws, one can again express all the
critical exponents in terms of the exponents $\sigma$ and $\tau$ this
time specifying the loop distribution.  

However, with the exception of random vortex tangles discussed in the
preceding subsection, usually not the percolation set of critical
exponents, but a different set is used to describe vortex loops.  The
difference can be seen by considering the correlation function in
percolation theory [above Eq.\ (\ref{algebraic})] and for random walks
[Eq.\ (\ref{G})], or more precisely their sum rules:
\begin{equation}
\label{sumGper} 
\sum_{\bf x} \; G_{\rm per}(x) \propto \sum_s s^2 \ell_s \propto \sum_s
s^{2- \tau } \, \mathrm{e}^{- c s},
\end{equation} 
and
\begin{equation} 
\label{sumG}
\sum_{\bf x} \; G(x) \propto \sum_n n^\tau \ell_n \propto \sum_n {\rm
e}^{- \beta \theta n a},
\end{equation} 
respectively, where the sum $\sum_{\bf x}$ is over all lattice sites.
The summands at the right sides of Eqs.\ (\ref{sumGper}) and
(\ref{sumG}) are seen to depend differently on the size ($s$ and $n$,
respectively).  To understand the origin of this difference, note that
the left sides equal the susceptibility $\chi$ for the respective order
parameters.  In general, $\chi$ is calculated from the partition
function by differentiating it twice with respect to the conjugate field
coupling linearly to the order parameter.  The conjugate field is, for
example, the magnetic field for magnetic systems, and the pressure for
liquid-gas transitions.  In other words, these correlation functions are
tied to the definition of the order parameter, each having its own set
of critical exponents.

For random walks, the set of critical exponents read in terms of the
exponents $\sigma$ and $\tau$ specifying the loop size distribution
\cite{NguyenSudbo,Hoveetal,perco}:
\begin{equation} 
\beta_{\rm ce} = \frac{\tau -2}{2\sigma}, \quad \gamma =
\frac{1}{\sigma}, \quad \eta = 2 - D.
\end{equation}
Here, the critical exponent $\beta$, which specifies the vanishing of
the order parameter when the critical temperature is approached, is
given a subscript ``ce'' (for critical exponent) to distinguish it from
the inverse temperature.  These formulas should be compared to the
analogous formulas in Eq.\ (\ref{pec}) for percolation.  The expressions
for $\gamma_{\rm per}$ and $\gamma$ are directly related to the
size-dependences in Eqs.\ (\ref{sumGper}) and (\ref{sumG}).  Brownian
random walks, having fractal dimension $D=2$, separate self-avoiding
vortex loops ($D<2$) from self-seeking ones ($D>2$).

\begin{figure}
\begin{center}
\includegraphics[width=4.0cm]{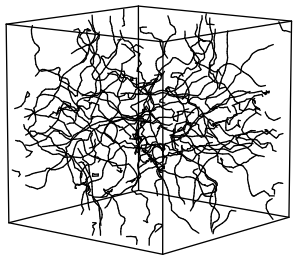} 
\end{center}
\caption{Computed vortex tangle in pure superfluid turbulence without
normal component.  From Araki, Tsubota, and Nemirovskii
\protect\cite{ATN}.
\label{fig:araki}}
\end{figure}

The two different sets of critical exponents, for percolation and for
random walks, both satisfy the scaling laws.  As only one correlation
length features in both descriptions, cf.\
Ref.~\onlinecite{WeinribTrugman}, the two sets have the exponent $\nu$
in common, implying that they are not independent.  Consider, for
example, the O(2), or XY-model, whose thermal phase transition admits a
description in terms of both correlated percolation and vortex loops.
Let $\sigma_{\rm per}$ and $\tau_{\rm per}$ specify the cluster size
distribution discussed in Sec.\ \ref{sec:corper}, and $\sigma$ and
$\tau$ the vortex loop size distribution (\ref{loopdis}), then
\begin{equation} 
\label{link}
\frac{\tau-1}{\tau_{\rm per}-1} = \frac{\sigma}{\sigma_{\rm per}},
\end{equation} 
or written in terms of fractal dimensions,
\begin{equation} 
\frac{D_{\rm per}}{D} = \frac{\sigma}{\sigma_{\rm per}},
\end{equation} 
where $D_{\rm per}$ denotes the fractal dimension of the (bond)
clusters, while $D$ denotes that of the vortex loops at criticality.

\subsection{Generating Vortex Tangles}
\label{sec:examples}
The vortex tangles in superfluids discussed so far were generated by
thermal fluctuations in the vicinity of the phase transition.  At least
two other ways to generate vortex tangles in superfluids are known.

(i) Consider setting up a counterflow, so that the superfluid and normal
components move with different velocities, or dragging an object such as
a grid through the superfluid.  When a critical velocity is reached, the
superfluid becomes turbulent through the formation of a random vortex
tangle made up of quantized vortex loops \cite{turbulence} (see
Fig.~\ref{fig:araki}).  For a constant, uniform counterflow, numerical
simulations by Schwarz \cite{Schwarz} showed that the vortex tangle can
be self-sustained.  In such \textit{counterflow turbulence}, the normal
component is possibly laminar and only the superfluid component
turbulent.  In \textit{superfluid grid turbulence}, on the other hand,
the superfluid and normal components are believed to become locked
together and behave like in classical turbulence, at least on scales
where the quantization of vortices is irrelevant.

\begin{figure}[b]
\begin{center}
\includegraphics[width=3.5cm]{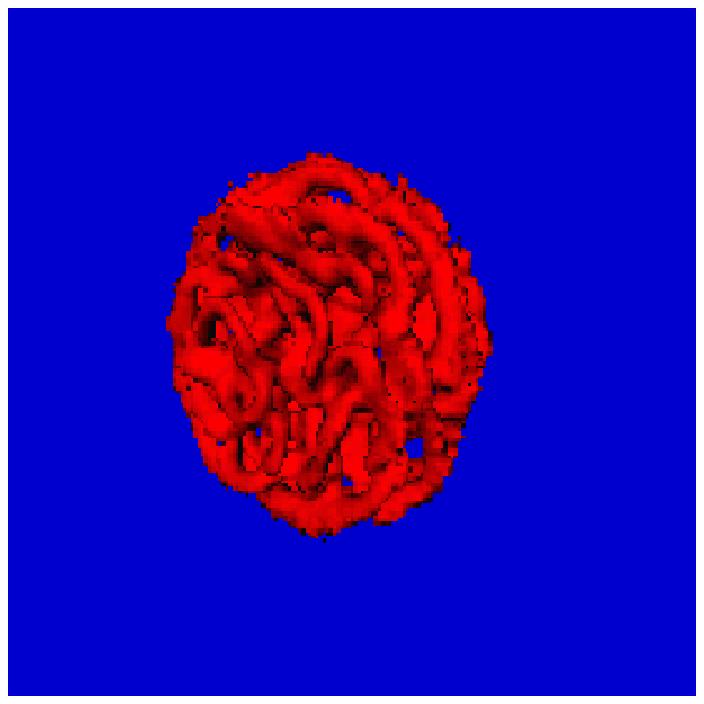}  \hspace{1cm}
\includegraphics[width=3.5cm]{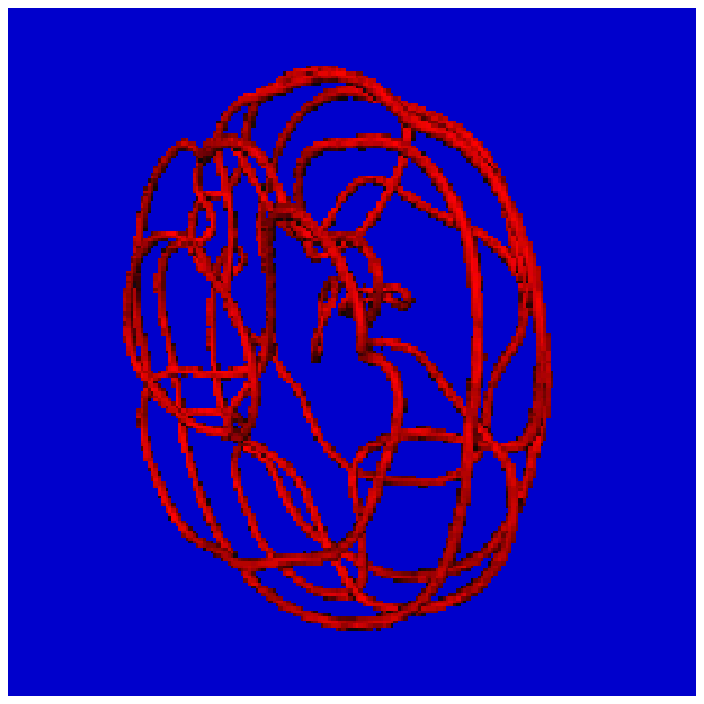} 
\end{center}
\caption{Computed evolution of a vortex tangle in superfluid $^3$He-B
after a rapid temperature quench of a heated bubble in the presence of a
superflow (perpendicular to the final configuration, resembling a snow
chain).  From Aranson, Kopnin, and Vinokur \protect\cite{AKV}.
\label{fig:kolia}}
\end{figure}

(ii) Consider heating a small region of the superfluid by, for example,
radiating the cell with neutrons as in the Helsinki experiment on
superfluid $^3$He-B in rotation \cite{Helsinki}.  When a neutron is
absorbed by the superfluid, the energy deposited causes a small region
to briefly heat up to the normal phase.  After the subsequent rapid
cool-down, a vortex tangle emerges.  This scenario for \textit{defect
formation} after a rapid phase transition was first proposed by Kibble
\cite{Kibble} in the context of cosmology and extended to condensed
matter by Zurek \cite{Zurek}.  From the present perspective, the
emergence of a vortex tangle after a rapid temperature quench is not so
much the formation of defects, as well as the ``freezing'' into
permanency of the dense vortex tangle rendering the hot bubble normal.
In this sense, the vortex tangle emerging after the quench is a remnant
of the normal phase.  In addition to infinite loops, the normal phase
has also many small vortex loops, which initially survive the rapid
temperature quench.  But as time passes, these small loops shrink to
zero, leaving behind only larger ones traversing a macroscopic portion
of the system (see Fig.~\ref{fig:kolia}).  The larger loops by no means
remain unchanged in time as they, through reconnections, spin off small
loops, or hook up with other vortex loops to grow.  The final
configuration depends, of course, on the conditions under which the
quench is performed, such as the presence of superflow.  Apart from
inflating vortex loops, which explains the expansion of the vortex
tangle in Fig.~\ref{fig:kolia}, a superflow can also lead to additional
mechanisms for defect formation, vortex loops wrapping around the bubble
being created to screen the superflow from the interior \cite{AKV}.

\section{Vortex Action}
\label{sec:va}
The discussion of vortex loops so far was qualitative as few details
are needed to describe phase transitions in terms of vortex
proliferation.  In this section, a more detailed description of vortex
lines is given.
\subsection{Schr\"odinger Equation}
\label{sec:schr}
To obtain the action describing vortex lines in a superfluid, the
differential equation (\ref{diff}) satisfied by the transition
probability $K_n({\bf x} \to {\bf x}')$ is written in the suggestive
form
\begin{equation} 
\label{Schr}
\partial_s K_s({\bf x} \to {\bf x}') = \tfrac{1}{2} \hbar \nabla^2
K_s({\bf x} \to {\bf x}')
\end{equation} 
by introducing the arclength parameter $s$,
\begin{equation} 
\label{s}
s = \frac{a^2}{d \hbar} n.
\end{equation} 
In this way, the differential equation becomes analogous to the
time-dependent Schr\"odinger equation for the ``time'' evolution
operator $K_s({\bf x} \to {\bf x}')$, satisfying
\begin{equation} 
\psi({\bf x}',s) = \int \mathrm{d}^d x \, K_s({\bf x} \to {\bf x}') \,
\psi({\bf x},0),
\end{equation} 
where $\psi({\bf x},s)$ is the wave function at position ${\bf x}$ and
``time'' $s$.  The arclength parameter does not represent real time, but
is the Schwinger propertime parameter \cite{propertime}.  It
parametrizes the path taken by the random walker, which is now pictured
as a (bosonic) particle trajectory.  The corresponding action reads:
\begin{equation} 
\label{S0}
S_0 = \sum_q \int \mathrm{d} s \left[ \tfrac{1}{2} \dot{\bf x}_q^2(s) +
\vartheta \right], \quad \vartheta = d \hbar^2 \beta \theta/a ,
\end{equation} 
where an extension to arbitrary many particles (labeled by $q$) is
given, and the second term resulting from the modification
(\ref{modification}) is included, with $\theta$ the vortex line tension.
The action is given a subscript 0 to indicate that the interaction
between vortex lines has been ignored for now.  With one of the spatial
coordinates interpreted as the time coordinate, the action (\ref{S0})
describes for $\theta \geq 0$ the worldlines of relativistic particles
of mass $\sqrt{2 \vartheta}$ in $d$-dimensional spacetime
\cite{Feynman50}.  The particles are at zero temperature, with the
quantum fluctuations representing the thermal fluctuations of the vortex
system at finite temperature.

To transcribe the (Euclidean) action (\ref{S0}), obtained using the
quantum-mechanical analog into an Hamiltonian, the dimensionless
combination $S/\hbar$ is to be replaced with $\beta H$, so that the
exponential factor $\exp(-S/\hbar)$ becomes a Boltzmann factor:
\begin{equation}
\label{analog} 
\exp(-S/\hbar) \to \exp(-\beta H).
\end{equation} 
The Planck constant $\hbar$ characterizing the quantum fluctuations
of the equivalent zero-temperature particle system is replaced with
the inverse temperature $\beta$ characterizing the thermal fluctuations
of the vortex system.  The transcription is achieved by the change of
variable
\begin{equation} 
s \to s' = \tfrac{1}{2} \hbar \beta s,
\end{equation} 
where the factor $\frac{1}{2}$ is included for mere convenience.  The
transcribed action then reads
\begin{equation} 
\label{H0}
H_0 = \sum_q \int \mathrm{d} s \left[ \tfrac{1}{4} \, \dot{\bf x}_q^2(s)
+ \epsilon_\mathrm{L}^2 \right], \quad \epsilon_\mathrm{L}^2 =
\frac{2d}{\beta} \frac{\theta}{a},
\end{equation} 
where the prime on the arclength parameter is suppressed, and the
parameter $\epsilon_\mathrm{L}$, which is determined by the vortex line
tension $\theta$, has the dimension of an energy per unit length.  It
follows from this expression that the line tension should scale linearly
with $a$ to obtain a finite value for the parameter
$\epsilon_\mathrm{L}$ in the continuum limit.  The correlation function
(\ref{G}) reads in these variables
\begin{equation} 
\label{G0}
G_0(|{\bf x} - {\bf x}'|) = \frac{2 d}{\beta a^2} \int_0^\infty
\mathrm{d} s \, K_s({\bf x} \to {\bf x}') = \frac{2 d}{\beta a^2}
\int_0^\infty \mathrm{d} s \, \left( \frac{\beta}{4 \pi s} \right)^{d/2}
\exp \left[- \frac{\beta}{4} \frac{({\bf x} - {\bf x}')^2}{s}\right] \,
\mathrm{e}^{- \beta \epsilon_\mathrm{L}^2 s}.
\end{equation} 
The exponent $d/2$ in the first factor of the integrand at the right
hand expresses the fact that the random walker can roam in $d$
dimensions, each giving a contribution $\sqrt{\beta/4 \pi s}$.

\subsection{Interaction}
\label{sec:inter}
\subsubsection{Superfluids}

\begin{figure}
\begin{center}
\psfrag{r}[t][t][1][0]{$R$} 
\includegraphics[width=7.0cm]{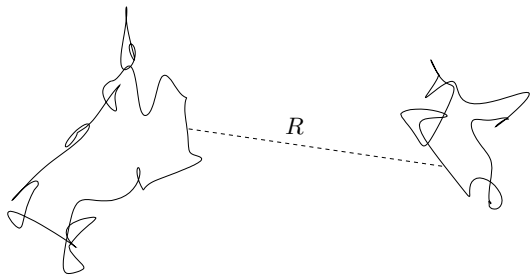} 
\end{center}
\caption{Interaction between two line segments belonging to two
different vortex loops.
\label{fig:inter}}
\end{figure}

In this subsection, the interaction between vortex loops in a superfluid
is considered.  The pairwise interaction is mediated by the Goldstone
mode associated with the spontaneously broken U(1) symmetry. Being
gapless, this mode leads to a $1/R$ interaction in three space
dimensions,
\begin{equation} 
\int \frac{\mathrm{d}^3 k}{(2  \pi)^3} \frac{\mathrm{e}^{\mathrm{i}
\mathbf{k} \cdot \mathbf{R}}}{k^2}  \sim \frac{1}{R}.
\end{equation} 
Each segment of a vortex line interacts via this potential with all
other line segments of the same vortex as well as of the other vortices
(see Fig.~\ref{fig:inter}).  The potential energy $V$ of a vortex
tangle is therefore of the form
\begin{equation} 
\label{V}
V = \frac{1}{2} \frac{g^2}{4 \pi} \sum_{q,q'} \oint \mathrm{d} s \oint
\mathrm{d} s' \, \dot{\bf x}_q (s) \cdot \frac{1}{R} \, \dot{\bf x}_{q'}
(s').
\end{equation} 
Here, $g$ is the charge of a vortex given by 
\begin{equation} 
\label{charge}
g^2 = \left(\frac{h}{m}\right)^2 \rho_{\rm s},
\end{equation} 
with $h/m$ the circulation quantum and $\rho_{\rm s}$ the superfluid
mass density, while $R$ is the distance between two line segments, $R =
|{\bf x}_q(s) - {\bf x}_{q'} (s')|$.  The mass parameter $m$ denotes the
mass of the atoms (for bosonic superfluids) or that of a pair of atoms
(for fermionic superfluids).  The integrations in Eq.\ (\ref{V}) are
along the vortex loops parametrized by their arclength parameters $s$
and $s'$.  Since $V$ decreases with increasing $R$, the pair interaction
between vortex segments with like-charges is repulsive.  The total
Hamiltonian $H$ describing a vortex tangle consists of the sum of the
free part (\ref{H0}) and the potential energy (\ref{V}),
\begin{equation} 
H = H_0 + V.
\end{equation} 

For two slightly deformed rectilinear vortices parallel to the $z$ axis, each
of length $L$ and separated by a distance $r(z)$, the right hand of Eq.\
(\ref{V}) reduces to the well-known logarithmic form
\begin{equation} 
V = \frac{g^2}{4 \pi} \int_0^L \mathrm{d} z \int_0^L \mathrm{d} z'
\frac{1}{\sqrt{r^2(z) + {z'}^2}} = - \frac{g^2}{4 \pi} \int_0^L
\mathrm{d} z \ln \left(\frac{r(z)}{L}\right).
\end{equation} 
The dependence on the vortex length $L$ in the argument of the logarithm
derives from the slow fall-off of the $1/R$ interaction.
\subsubsection{Superconductors}
\begin{figure}
\begin{center}
\psfrag{z}[t][t][1][0]{$z$} 
\psfrag{h}[t][t][1][0]{$h_\mathrm{ext}$} 
\includegraphics[width=12.0cm]{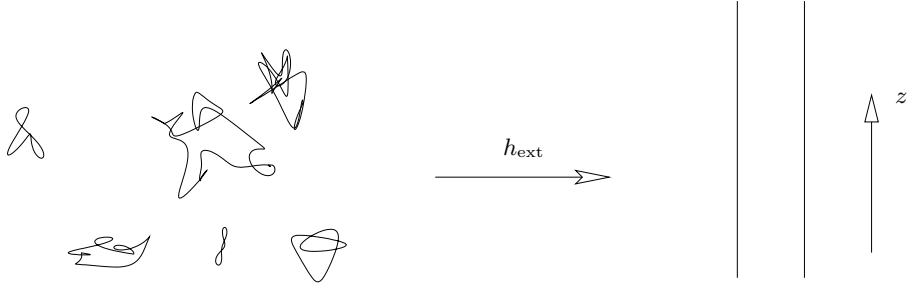} 
\end{center}
\caption{Suppression of vortex loops and creation of open vortex lines
when an external magnetic field is applied. 
\label{fig:nr}}
\end{figure}

The results obtained for superfluids are easily adapted for
superconductors.  The major difference is that the would-be Goldstone
mode turns the long-ranged electromagnetic interaction into a screened
one through the Higgs mechanism.  This leads to the modification
\begin{equation} 
\frac{1}{R} \to \frac{\mathrm{e}^{-R/\lambda}}{R}
\end{equation} 
in Eq.\ (\ref{V}), where $\lambda$ is the London penetration depth.  The
charge $g$ of a magnetic vortex is again given by Eq.\ (\ref{charge})
for a superfluid, although it is usually put in the equivalent form
\begin{equation} 
\label{chargem}
g = \frac{\Phi_0}{\lambda},
\end{equation} 
where $\Phi_0$ is the magnetic flux quantum.  Note that $g$ is
independent of the electric charge $e$.  The $e$-dependences in $\Phi_0$
and the penetration depth precisely cancel in the right hand of Eq.\
(\ref{chargem}).  Since the interaction is screened, the potential
energy between two parallel slightly deformed rectilinear magnetic
vortices separated by a distance $r(z)$ is
\begin{equation} 
\label{intersc}
V = \frac{g^2}{4 \pi} \int_0^L \mathrm{d} z \int_0^L \mathrm{d} z'
\frac{\mathrm{e}^{-\sqrt{r^2(z) + {z'}^2}/\lambda}}{\sqrt{r^2(z) +
{z'}^2}} = \frac{g^2}{4 \pi} \int_0^L \mathrm{d} z \, K_0
\left(\frac{r(z)}{\lambda}\right),
\end{equation} 
with $K_0$ a modified Bessel function of the second kind, which for
small argument $x$ behaves as $K_0(x) \sim -\ln(x)$.  The role of
infrared cutoff is now played by the London penetration depth $\lambda$.

\subsection{External Field}
\label{sec:exfield}

From the present perspective, the major effect of turning on an external
magnetic field is to suppress fluctuations along the field.  In other
words, the random walker representing the vortex trajectory is forced to
move only in the direction of the applied field, and cannot move
backwards (see Fig.~\ref{fig:nr}).  The external field thus suppresses
the formation of vortex loops and (under certain conditions discussed
below) instead leads to open vortices along the field direction,
starting at the bottom of the sample and terminating at the top. The
vortices can now only fluctuate around their equilibrium position by
making excursions in the plane perpendicular to the field.  In the
quantum-mechanical analog, this corresponds to taking the
nonrelativistic limit.

\begin{figure}
\begin{center}
\psfrag{z}[t][t][1][0]{$\mathrm{d}z$} 
\psfrag{p}[t][t][1][0]{$\mathrm{d}z'$} 
\includegraphics[width=1.0cm]{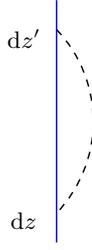}
\end{center}
\caption{Two line segments of the same vortex (solid line) interacting
via the $1/R$ potential (dashed line). \label{fig:single}}
\end{figure}

To study the effect of applying an external field, which is assumed to
be parallel to the $z$ axis, write
\begin{equation} 
\label{specrel}
({\bf x} - {\bf x}')^2 = {\bf r}^2 + c^2 z^2,
\end{equation} 
where ${\bf r}(z) = [x(z),y(z)]$ are the coordinates perpendicular to
the field, parametrized by $z$.  The parameter $c$ is introduced (via
the replacement $z \to c z$) to facilitate interpreting the spatial
coordinate $z$ as a time coordinate.  The observation that fluctuations
along the field are suppressed is implemented by letting $c \to \infty$.
In this limit, the integral over the Schwinger propertime parameter $s$
in Eq.\ (\ref{G0}) can be approximated by the saddle point
\cite{Stephens}
\begin{equation} 
\label{sz}
s = \frac{c}{2 \epsilon_\mathrm{L}} z,
\end{equation} 
connecting $s$ to the timelike variable $z$, with the result in three
dimensions
\begin{equation} 
\frac{\beta a^2}{2 d} G_0(|{\bf x} - {\bf x}'|) \approx
\left(\frac{\beta}{4 \pi z} \right)^{2/2}\, \mathrm{e}^{ - \beta
\epsilon_\mathrm{L} (z + {\bf r}^2/2 z)} .
\end{equation}
Here, the parameter $c$ is scaled away again by the replacement $c z \to
z$ thereby undoing the earlier replacement in Eq.\ (\ref{specrel}). The
exponent in the prefactor at the right hand is unity now instead of
$3/2$ as it was in the absence of a field.  It indicates that after
applying the magnetic field only the two directions perpendicular to the
field are still available to the random walker to roam, each giving a
contribution $\sqrt{\beta/4 \pi z}$.  The Hamiltonian (\ref{H0}) becomes
in this limit
\begin{equation} 
\label{H0nr}
H_0 = \sum_q \int \mathrm{d} s \left[ \tfrac{1}{4} \, \dot{\bf x}_q^2(s)
+ \epsilon_\mathrm{L}^2 \right] \to \sum_q \int \mathrm{d} z \left[
\tfrac{1}{2} \, \epsilon_\mathrm{L} \dot{\bf r}_q^2(z) +
\epsilon_\mathrm{L} \right],
\end{equation} 
where use is made of Eq.\ (\ref{sz}).  This procedure of taking the
nonrelativistic limit is equivalent to the one used by Fetter
\cite{Fetter} of expanding in small displacements in the $xy$ plane.
With $z$ interpreted as the time coordinate, this Hamiltonian becomes,
using the transcription (\ref{analog}), an action describing
nonrelativistic bosons of mass $\epsilon_\mathrm{L}$ moving in a plane
in the presence of a constant background potential
$\epsilon_\mathrm{L}$.  As in the absence of an external field, the
vortices, which are now open lines, are interpreted as the worldlines of
particles, with the quantum fluctuations representing the thermal
fluctuations of the vortex system.  The limit taken in Eq.\ (\ref{H0nr})
precisely corresponds to the nonrelativistic limit of the relativistic
Hamiltonian (\ref{H0}).  The interaction between two ``nonrelativistic''
vortex lines is given by Eq.\ (\ref{intersc}) as the case of slightly
deformed rectilinear vortices considered there precisely describes
vortices in the presence of an external field.

Not only two line segments belonging to different vortices interact via
the $1/R$ potential, but also two line segments of the same vortex (see
Fig.~\ref{fig:single}), giving the contribution to the energy
\begin{equation} 
\label{Abri}
\frac{g^2}{4 \pi} \int_0^L \mathrm{d} z \int_0^L \mathrm{d} z' \,
\frac{{\rm e}^{-\sqrt{(z - {z'})^2}/\lambda}}{\sqrt{(z - {z'})^2}} =
\frac{g^2}{4 \pi} \int_0^L \mathrm{d} z \, \ln
\left(\frac{\lambda}{\xi}\right),
\end{equation} 
where the coherence length $\xi$ is taken as ultraviolet cutoff.  This
contribution amounts to a renormalization of the parameter
$\epsilon_\mathrm{L}$.  In the following, that parameter, introduced in
Eq.\ (\ref{H0}), is assumed to be given by the integrand at the right
hand of Eq.\ (\ref{Abri}), representing the self-interaction energy per
unit length, i.e.,
\begin{equation} 
\epsilon_\mathrm{L} \to \frac{g^2}{4 \pi} \ln (\kappa),
\end{equation} 
where $\kappa =\lambda/\xi$ is the Ginzburg-Landau parameter.  The total
Hamiltonian $H$ describing vortices in the presence of an applied
magnetic field is then given by the sum of the free part (\ref{H0nr})
and the interaction part
\begin{equation} 
\label{Hintnr}
H_\mathrm{int} = \frac{g^2}{4 \pi} \sum_{q \neq q'} \int \mathrm{d} z \,
K_0 \left(\frac{|{\bf r}_q (z) - {\bf r}_{q'} (z)|}{\lambda}\right) -
\Phi_0 \sum_q \int \mathrm{d} z \, h_\mathrm{ext}[{\bf r}_q (z)],
\end{equation} 
where the last term represents the interaction of the vortex lines with
the external field.  This field penetrates the sample only if the sum of
the constant terms in Eqs.\ (\ref{H0nr}) and (\ref{Hintnr}) becomes
negative, i.e., when
\begin{equation} 
h_{\rm ext} > \frac{\epsilon_\mathrm{L}}{\Phi_0} = h_{{\rm c}_1},
\end{equation} 
defining the lower critical field $h_{{\rm c}_1}$.  Below this critical
field, the system is in the Meissner state where the external field is
excluded from the superconductor.  Above $h_{{\rm c}_1}$, the system is
in the mixed state where the external field invades the superconductor
through an array of magnetic vortex lines.  When fluctuations can be
ignored, the repulsive pair interaction between the vortices makes them
form a triangular Abrikosov flux lattice \cite{Abrikosov}.

\section{Entangled Vortex Lines}
\label{sec:entangled}
In this section, vortex tangles of a different kind than the ones above
are discussed.  Instead of consisting of closed vortex loops, the
tangles considered here consist of open vortex lines, connecting the top
and bottom of the sample due to an applied magnetic field (for
superconductors) or rotation (for superfluids).
\subsection{Vortex Lattice Melting}
\label{sec:melting}
\begin{figure}
\begin{center}
\includegraphics[width=7cm]{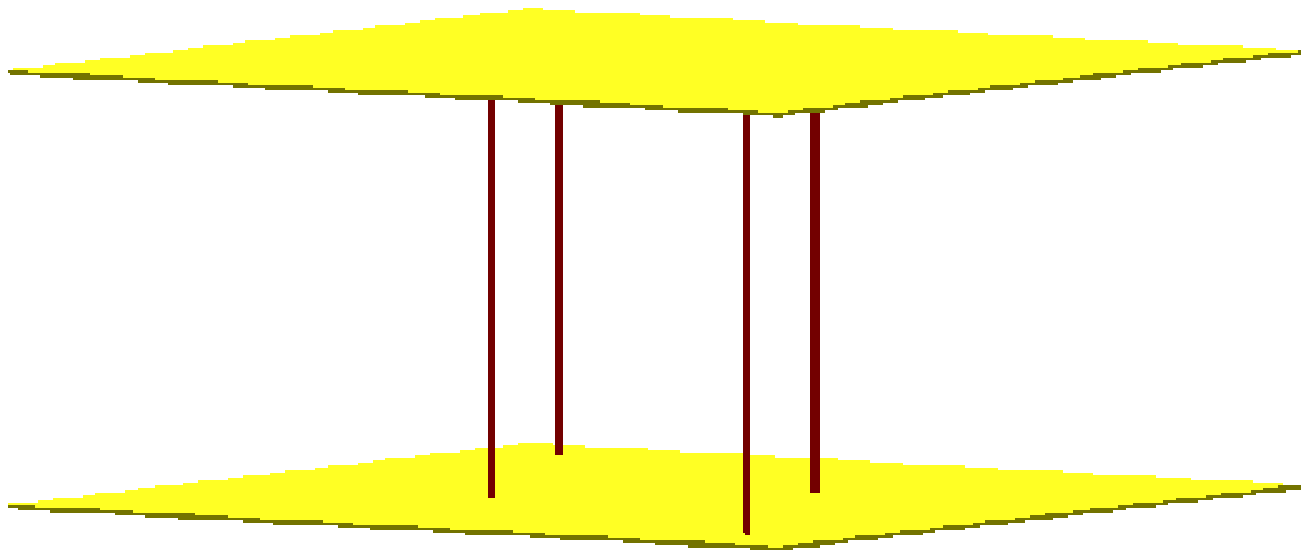} 
\hspace{1cm}
\includegraphics[width=7cm]{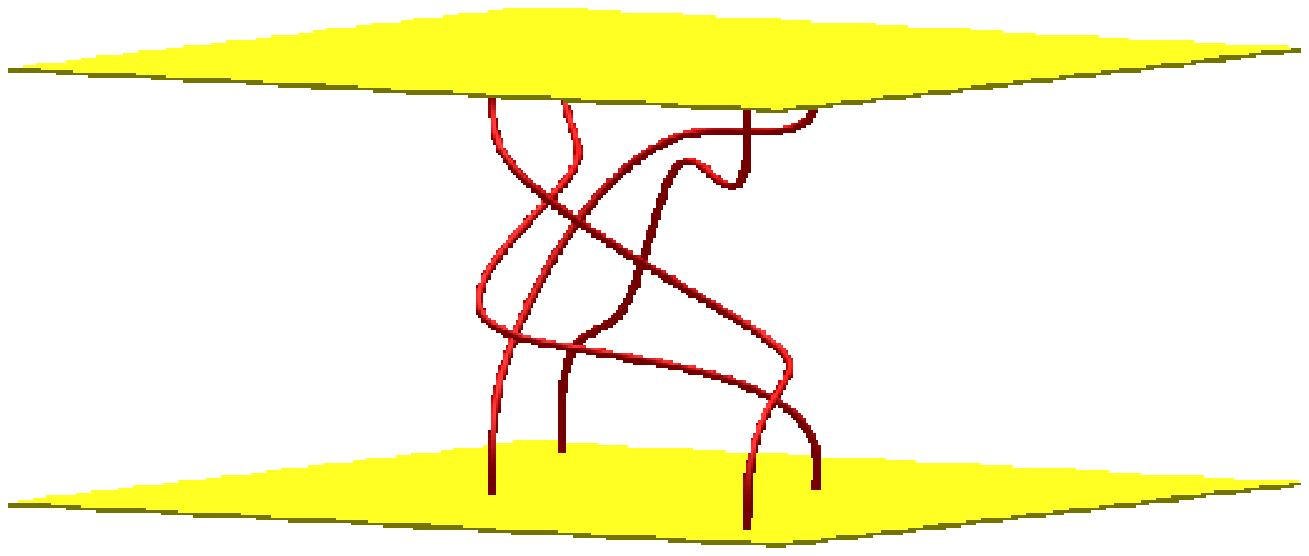}
\end{center}
\caption{Artist's impression of the Abrikosov flux lattice (left panel)
and the entangled vortex liquid (right panel).  Periodic boundary
conditions in the $z$ direction are assumed. \label{fig:nonrel}}
\end{figure}

The equivalent particle system of nonrelativistic interacting bosons in
2+1 dimensions (with the $z$ coordinate interpreted as a timelike
variable) specified by the Hamiltonian $H = H_0 + H_\mathrm{int}$ given
in Eqs.\ (\ref{H0nr}) and (\ref{Hintnr}) was taken by Nelson
\cite{Nelson} as starting point to investigate the effect of thermal
fluctuations on the Abrikosov vortex lattice.  A similar approach to
study fluctuations in vortex lattices in rotating $^4$He and in
superconductors was pioneered by Fetter \cite{Fetter}.  In the particle
system, the vortex lattice corresponds to a Wigner crystal resulting
from the repulsive pairwise interaction between the particles.  The
thermal fluctuations of the vortex system translate into quantum
fluctuations of the equivalent particle system, which, when violent
enough, were predicted to melt the crystal.  The resulting phase in the
equivalent particle system is the superfluid phase \cite{Nelson}, which
is generally believed to be the sole alternative to the crystal phase
for interacting bosons at low temperatures in a clean system without
impurities.  The superfluid phase corresponds to entangled vortex lines
in the vortex system (see Fig.~\ref{fig:nonrel}).

The importance of quantum fluctuations in a system of interacting bosons
is measured by the so-called de Boer parameter $\Lambda$ \cite{deBoer}.
This dimensionless parameter can be obtained by writing the kinetic term
at the right hand of Eq.\ (\ref{H0nr}) and the pair interaction term in
Eq.\ (\ref{Hintnr}) as
\begin{equation} 
\beta H' = \int \mathrm{d} u \left[ \sum_q \frac{1}{2 \Lambda^2} \,
\dot{\bf r}_q^2(u) + \sum_{q < q'} K_0 \left(\frac{|{\bf r}_q (u)
- {\bf r}_{q'} (u)|}{\lambda}\right) \right],
\end{equation} 
where $u = \beta a g^2 z/2 \pi$ and 
\begin{equation} 
\frac{1}{\Lambda} = \beta a \sqrt{g^2 \epsilon_\mathrm{L}/2\pi},
\end{equation} 
with all length scales now being given in units of a typical length $a$,
e.g., the lattice spacing when working on a lattice.  The vortex lattice
melting transition into an entangled vortex liquid was numerically shown
to take place at the critical value $\Lambda_\mathrm{m} \sim 0.062$
\cite{MaCe}.  There is ample numerical \cite{numerical} as well as
experimental \cite{experimental} evidence that the transition is
discontinuous in a clean system.  Further experimental results have been
reported \cite{abcp}, indicating that the discontinuous melting line in
the $h$-$T$ phase diagram terminates at a critical point in the
high-field, low-temperature corner.

\subsection{Feynman's cooperative exchange rings}
\label{sec:rings}
\begin{figure}
\begin{center}
\psfrag{y}[t][t][1][0]{$y,z$} 
\psfrag{x}[t][t][1][0]{$x$}
\includegraphics[width=5cm]{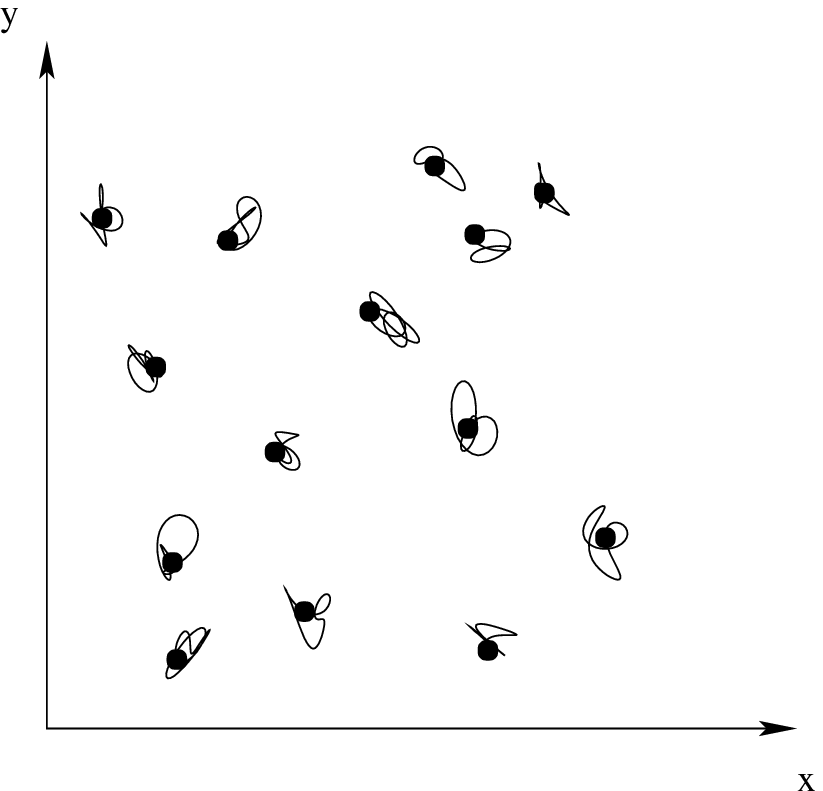} \hspace{2cm}
\includegraphics[width=5.cm]{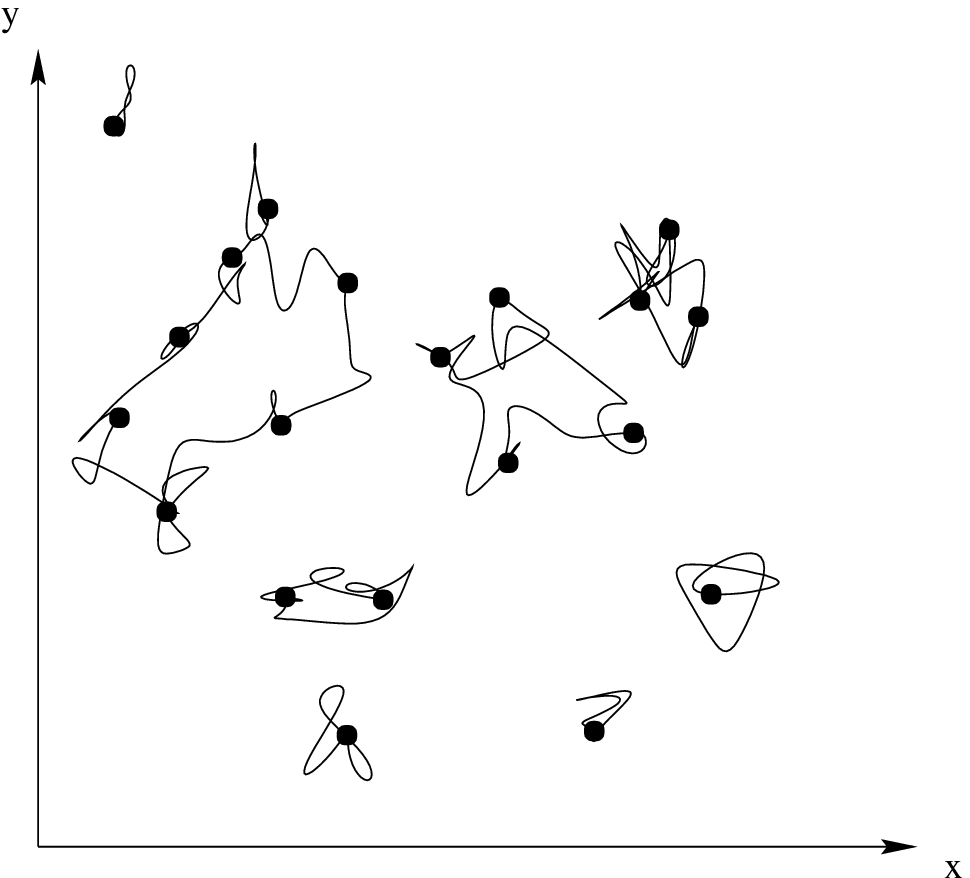} 
\end{center}
\caption{Trajectories of \textit{non-interacting} bosons in three space
dimensions executed during the (imaginary) time interval $\hbar \beta$.
For ease of representation the $y$ and $z$ axes are lumped together.
The starting points of the Brownian random walks are indicated by dots.
In the normal phase (left panel), most random walks end at their own
starting point, whereas in the Bose-Einstein condensed phase (right
panel) large cooperative exchange rings are formed, with the trajectory
of one particle ending at the starting point of another.  In this way,
particles in a ring are cyclically permuted after an imaginary time
$\hbar\beta$.  (Adapted from Ref.~\protect\onlinecite{loops}.)
\label{fig:dots}}
\end{figure}

As was pointed out by Nelson \cite{Nelson}, the entangled vortex liquid
can be elegantly understood in the equivalent particle system, using
Feynman's cooperative exchange ring theory of the $\lambda$ transition
in $^4$He \cite{lambda}.  Bose-Einstein condensation (BEC) is understood
in this picture as follows.  At finite temperature, the
(3+1)-dimensional $^4$He particle trajectories form closed loops because
the (Euclidean) time coordinate is compactified, taking values only in
the interval $[0, \hbar \beta]$.  For an ensemble of bosons, the
boundary conditions are periodic, meaning that the configuration at time
$0$ and at time $\hbar \beta$ are identical.  In the normal phase, most
bosons execute during this time interval a random walk in space (see
left panel of Fig.~\ref{fig:dots}), which starts and ends at the same
position (indicated by a dot).  It means that the particles, being
distinguishable, behave classically.  When the critical temperature is
approached from above, the particles lose their identity and become
indistinguishable.  In Feynman's theory this is reflected by the
formation of so-called cooperative exchange rings, where individual
worldlines hook up to form larger loops \cite{Ceperley} (see right panel
of Fig.~\ref{fig:dots}).  A particle in such a composite ring moves in
imaginary time along a trajectory that does not end at its own starting
position, but at that of another particle.  Hence, although the initial
and final configurations are identical, the particles in a composite
ring are cyclically permuted.

To determine the condition for this to happen, consider an ideal Bose
gas for simplicity.  The non-interacting particles are described by a
time evolution operator $K_\tau({\bf x} \to {\bf x}')$, satisfying the
time-dependent Schr\"odinger equation (\ref{Schr}) with the Schwinger
propertime parameter $s$ introduced in Eq.\ (\ref{s}) now replaced by
the (Euclidean) time
\begin{equation} 
\tau =  \frac{a^2m}{d \hbar} n.
\end{equation} 
The right hand explicitly depends on the particle mass $m$.  It follows
that during the time interval $\Delta \tau = \hbar \beta$, the particles
execute a Brownian random walk with an average length
\begin{equation} 
\label{Lav}
L_{\hbar \beta} = \frac{d}{2 \pi} \frac{\lambda_\beta^2}{a}
\end{equation} 
determined by the de Broglie thermal wavelength $\lambda_\beta$.  If the
length scale $L_{\hbar \beta}$ becomes, upon reducing the temperature,
on the order of the interparticle distance, individual worldlines can
hook up to form large loops.  The particles participating in the
exchange rings become indistinguishable and the system Bose-Einstein
condenses.  When worked out, this condition gives an estimate of the
critical temperature $T_{\rm c}$.

A ring made up of the worldlines of $w$ particles wraps around the
imaginary time cylinder $w$ times.  BEC corresponds in this picture to
the appearance of long loops, wrapping arbitrarily many times around the
imaginary time cylinder \cite{Stone}.  In other words, as for vortex
loops before (see Sec.\ \ref{sec:RW}), this transition corresponds to a
proliferation of loops---not of vortex loops, but of worldline loops
this time.  BEC can therefore be described as a percolation process
\cite{Suto}.  The worldline loop size distribution \cite{loops}
\begin{equation}
\label{ellw}
\ell_w  \propto w^{- \tau} \mathrm{e}^{-(\alpha - \alpha_{\rm c}) w},
\end{equation} 
with $\alpha = -\beta \mu$ and $\tau = d/D+1$, gives the density of rings
containing $w$ particles, or, equivalently, the density of loops
wrapping around the imaginary time cylinder $w$ times.  The ``line
tension'' $\alpha - \alpha_{\rm c}$ of the worldlines is determined by
the chemical potential $\mu$ [as was the Boltzmann factor in the cluster
size distribution (\ref{abbr}) in Mayer's theory of the liquid-gas
transition] and vanishes when the critical temperature is approached
from above [cf.\ Eq.\ (\ref{dislg})].  Specifically,
\begin{equation} 
\alpha - \alpha_{\rm c} \propto (T - T_{\rm c})^{1/\sigma}.
\end{equation} 
The worldline loop size distribution then changes from an exponential
decay with increasing loop length to an algebraic decay.  The winding
number $w$ in Eq.\ (\ref{ellw}), denoting the number of particles
contained in a ring plays the role of the number $n$ of steps in the
loop size distribution (\ref{loopdis}).  In other words, a straight path
segment traversed during a single step by a classical random walker now
corresponds to a random trajectory traversed by a quantum particle
during the time interval $\hbar \beta$ [with an average length $L_{\hbar
\beta}$ given by Eq.\ (\ref{Lav}) if the particle is non-interacting].
Since the loop size distribution is of statistical nature, details which
require a resolution greater than the step size are irrelevant---whence
the similarity in the loop distributions (\ref{ellw}) and
(\ref{loopdis}).  Given the two exponents $\sigma$ and $\tau$ specifying
the worldline loop size distribution (\ref{ellw}), the critical
exponents characterizing the phase transition can again be obtained
using scaling laws \cite{perco}.

Feynman's picture of the $\lambda$ transition is dual to the one based
on vortex proliferation discussed earlier.  Depending on whether the
critical temperature is approached from below or above, vortices (in the
superfluid phase), or worldlines (in the normal phase) proliferate.
This is similar to the duality discussed in the context of percolation
at the end of Sec.\ \ref{sec:up}, where clusters of the minority phase
are considered in the sea made up of the majority phase: here either
vortices in a sea of proliferated worldlines, or worldlines in a sea of
proliferated vortices are considered.

Translated back to the vortex system, Fig.~\ref{fig:dots} represents
the projections of the vortex lines onto a plane perpendicular to the
applied field.  The periodic boundary conditions of the equivalent
particle system are somewhat unnatural for the vortex system, but are
believed not to dramatically change the results, provided $L$, playing
the role of $\hbar \beta$ in the equivalent particle system, is large
enough.  The left panel of Fig.~\ref{fig:dots} mimics the Abrikosov
flux lattice (apart from the lattice symmetry), with the vortex lines
making only small excursions from the straight lines connecting the
endpoints, while the right panel mimics the entangled vortex liquid.
Precisely such snapshots of vortex positions were recently obtained in
numerical simulations by Sen, Trivedi, and Ceperley \cite{numerical}.

\subsection{Order Parameter of Entangled Vortex Liquid}
\label{sec:op}
A superfluid phase is characterized by the superfluid mass density
$\rho_{\rm s}$ giving the response of the system to an externally
imposed velocity ${\bf v}$.  A computationally simple, yet powerful way
of obtaining the superfluid mass density on a finite lattice with
periodic boundary conditions was introduced by Pollock and Ceperley
\cite{PollockCeperley}, who related $\rho_{\rm s}$ to the winding of
particle trajectories around the lattice (not in the time direction, but
in the space directions).

The free energy $F = - \beta^{-1} \ln(Z)$ acquires an additional
term after the boost
\begin{equation} 
\Delta F_v = \tfrac{1}{2} \mathbb{V} \rho_{\rm s} {\bf v}^2.
\end{equation} 
In the equivalent particle picture, this boost is implemented by letting
\begin{equation} 
\tfrac{1}{2} m \dot{\bf r}_q^2 \to \tfrac{1}{2} m (\dot{\bf r}_q - {\bf
v})^2 \approx \tfrac{1}{2} m \dot{\bf r}_q^2 - m
\frac{\mathrm{d} {\bf r}_q}{\mathrm{d} \tau} \cdot {\bf v},
\end{equation} 
leading to 
\begin{equation}  
\label{equi}
{\rm e}^{- \beta \Delta F_v} = \left\langle \, \exp \left( {\rm i}
\frac{m}{\hbar} {\bf w} \cdot {\bf v} \right) \right\rangle ,
\end{equation}  
with the winding vector:
\begin{equation} 
{\bf w} = \sum_q \int_0^{\hbar \beta} \mathrm{d} \tau \frac{\mathrm{d}
{\bf r}_q(\tau)}{\mathrm{d} \tau}.
\end{equation} 
To understand the appearance of the imaginary unit at the right hand of
Eq.\ (\ref{equi}), note that in real time, particles are described by
the phase factor $\exp(\mathrm{i} S/\hbar)$.  When going over to the
Euclidean time variable $\tau$, the term linear in the velocity retains
the imaginary unit in the phase factor.

Since the integrand in the definition of the winding vector is a total
derivative, only the endpoints of the integration contribute.  In
numerical simulations on a finite lattice, usually also periodic
boundary conditions in the spatial directions are imposed.  The winding
vector then becomes
\begin{equation} 
\label{w}
{\bf w} = \sum_q \left[{\bf r}_q(\hbar \beta) - {\bf r}_q(0) \right] \,
+ \, L {\bf n} ,
\end{equation} 
with $L$ the linear size of space and ${\bf n}$ a $d$-component vector
with integer components, denoting the number of times the trajectories
wrap around space in the $x$, $y$, $\cdots$ direction, respectively.
Because of the periodic boundary condition in the $\tau$ direction, the
sum in Eq.\ (\ref{w}) is always zero, even when particles are cyclically
permuted, so that a nonzero value for the winding vector obtains only
when worldlines wrap around space, represented by the last term in in
Eq.\ (\ref{w}).  In the language used in connection with the length scale
(\ref{lc}), such loops are classified as infinite.  Expanding both sides
of Eq.\ (\ref{equi}) to quadratic order in the applied velocity ${\bf
v}$, one finds that \cite{PollockCeperley}
\begin{equation} 
\langle {\bf w}^2 \rangle = d \mathbb{V} \beta g^2,
\end{equation} 
with $g$ the vortex charge introduced in Eq.\ (\ref{charge}), $g^2 =
(h/m)^2 \rho_{\rm s}$.  It shows that the superfluid mass density is
nonzero when infinite worldline loops are present.  That is, $\rho_{\rm
s}$ directly signals the proliferation of these loops and is therefore
the proper order parameter describing the entangled vortex liquid.

The numerical simulations by Nordborg and Blatter \cite{numerical} show
that the superfluid mass density sharply rises at the critical value
$\Lambda_\mathrm{m} \sim 0.062$ from $\rho_{\rm s} \approx 0$ in the
crystal phase to $\rho_{\rm s} \approx \rho$ in the superfluid phase,
where $\rho$ is the total mass density. 

\section{Conclusion}
\label{sec:conclusion}
In these lecture notes, thermal phase transitions taking place in
superfluids were discussed from a geometrical perspective.  Two
seemingly different geometrical approaches were treated: one modeled
after cluster percolation and the other after loop proliferation first
proposed by Onsager \cite{Onsager}.  Apart from peculiarities resulting
from differences in definition, the two geometrical approaches were,
however, seen to be two faces of the same underlying formalism.  Central
in the description was the size distribution of clusters and loops,
specified by two exponents from which in turn all the critical exponents
characterizing the superfluid phase transition follow through scaling
relations.  At the critical point, two different types of clusters
percolate and also two different types of loops proliferate, depending
from which side the transition temperature is approached.  The
percolating cluster always belongs to the minority phase which then
turns into the majority phase.  The two percolation descriptions are
dual to each other in that both phases are interchanged.  Also the two
loop descriptions are dual to each other, with finite vortex loops
featuring in the superfluid phase and finite worldline loops featuring
in the normal phase.  These finite loops proliferate when approaching
the critical temperature after which their roles are interchanged.

Finally, the similarity between the cluster distribution (\ref{dislg})
and the worldline loop distribution (\ref{ellw}), both depending on the
fugacity $\exp(-\alpha)$, indicates that the descriptions in terms of
clusters and worldline loops are in fact closely related.  This
connection is reinforced by the surprising presence of the de Broglie
wavelength (which depends on $\hbar$) in the classical partition
function (\ref{Zlg}).

\begin{acknowledgments}
I'm indebted to M. Krusius for the kind hospitality at the Low
Temperature Laboratory in Helsinki and for inviting me to lecture at the
Kevo Winter School, Kevo, Finland, 20-26 April 2002.  I wish to thank
H. Alles for organizing and running the pleasant and inspiring Winter
School; S. Balibar, R. Blaauwgeers, A. Fetter, R. H\"anninen, N. Kopnin,
M. Krusius, L. Skrbek, E. Thuneberg, J. Viljas, and G. Volovik for
fruitful discussions; T. Araki and N. Kopnin for allowance to reproduce
their Fig.~\ref{fig:araki} and \ref{fig:kolia}, respectively; and
R. Blaauwgeers for Figs.~\ref{fig:cut_ball} and \ref{fig:nonrel}.

This work was funded by the European Union program Improving Human
Research Potential (ULTI III).
\end{acknowledgments}

\end{document}